# LOFAR observations of asymmetric quasi-periodic scintillations in the mid-latitude ionosphere.

G. Dorrian[1], D. Themens[1], T. Renkwitz[2], G. Nykiel[3], A. Wood[1], B. Boyde[1], R. A. Fallows[4], M. Mevius[5], H. Trigg[1]

[1]Space Environment & Radio Science Group (SERENE), University of Birmingham, UK

[2]Liebniz Institute of Atmospheric Physics at the University of Rostock, Schloßstraße 6, 18225 Kühlungsborn, Germany

[3]Faculty of Civil and Environmental Engineering, Gdańsk University of Technology, 80-233 Gdańsk, Poland

[4]RAL Space, Science & Technology Facilities Council, Harwell Campus, Oxford, UK

[5]ASTRON, the Netherlands Institute for Radio Astronomy, Postbus 2, 7990 AA, Dwingeloo, The Netherlands

## Abstract

The LOw Frequency ARray (LOFAR) was used to track the propagation of a TID containing embedded plasma structures which generated type 1 asymmetric quasi-periodic scintillations (QPS: Maruyama, 1991) over a distance of >1200 km across Northern Europe. Broadband trans-ionospheric radio scintillation observations of these phenomena are, to our knowledge, unreported in the literature as is the ability to track asymmetric QPS generating plasma structures over such a distance. Type 1 asymmetric QPS are characterised by an initial broadband signal fade and enhancement which is then followed by 'ringing pattern' interference fringes. These are caused by diffractive fringing as the radio signal transitions through regions of relatively steep plasma density gradient at the trailing edge of the plasma structures. That the QPS retained their characteristics consistently over the full observing window implies that the plasma structures generating them likewise held their form for several hours, and over the full 1200 km distance. The most likely TID propagation altitude of 110 km was consistent with a persistent and non-blanketing sporadic-E region detected by the Juliusruh ionosondes, and direct measurements from co-located medium frequency radar. Co-temporal GNSS data was used to establish that these plasma density variations were very small, with a maximum likely amplitude of no more than +/- 0.02 TECu deviation from the background average. The observations were made between 0430-0800 UT on 17 December 2018 under very quiet geophysical conditions which possibly indicated a terrestrial source. Given the TID propagation direction, the source was likely located at high-latitude.

# 1. Introduction

Quasi-periodic scintillations (QPS) in the ionosphere are repeating fluctuations in the received power of backscattered VHF radar or trans-ionospheric radio observations which were first reported by Elkins & Slack (1969). In the past they have been referred to as 'ringing irregularities' (Doan & Forsyth, 1978) which is an apt term as QPS are often characterised by a large signal fade which can last several minutes, symmetrically bounded by decreasing amplitude enhancement fringes akin to the damped oscillations of a bell after it has been struck. In other cases they are asymmetric, with a signal fade either followed or preceded by decreasing secondary amplitude fringes (Maruyama, 1991).

Sporadic E-layers (Es) are thin layers of metallic ions and free electrons in the E-region ionosphere which are generally regarded to form by vertical wind shear in the neutral atmosphere or electric fields. The wind shear forces plasma along inclined field lines to concentrate into a thin layer a few km thick, or less (Kopp, 1997; Whitehead, 1961; Axford, 1963). It has also been demonstrated recently by Qiu et al., (2023) that Es formation can be influenced by the ratio of vertical gradient of ion-neutral collision frequency, and ion gyro-frequency.

Statistical observations by Hajkowicz & Dearden (1988) established that occurrence rates of QPS tend to peak in late morning (~0800-1000 LT) and pre-midnight (~2000-2200 LT) during summer. They also observed a ~40% decrease in occurrence rate at solar minimum with respect to solar maximum. Furthermore, they associated constant height sporadic-E with daytime QPS, and range-type spread-F with examples observed at night. Maruyama (1995) used a variety of plasma irregularity models, varying scale size, electron density, and morphology, to recreate observed examples of scattering and found that 82% of daytime QPS are caused by drifting 2-dimensional disk shaped plasma irregularities with a steep density gradient on the trailing edge, while approximately half of night time examples are caused by linear features. Radar observations by Maruyama et al., (2000) further examined the vertical spatial scale of these plasma irregularities and established that they can extend in thickness for several tens of kilometres, in contrast to nominally thin sporadic E-layers. More recently Birch & Hargreaves (2020) identified QPS in the high-latitude F-region using observations with the EISCAT Svalbard Radar, with a median inter-peak periodicity of between 20-26 minutes. These authors also linked the QPS with ripples detected in the magnetosphere and the solar wind.

Atmospheric gravity waves in the thermosphere manifest as travelling ionospheric disturbances (TIDs: Hines, 1960; Hooke, 1968; Hocke & Schlegel, 1996; Azeem et al., 2017), which can be detected by a variety of methods including their impact on trans-ionospheric radio propagation. Medium-scale TIDs (MSTID) in the mid-latitude ionosphere typically exhibit wavelengths of a few hundred km and propagate with subsonic phase velocities of 50-150 ms$^{-1}$ (e.g. Shiokawa et al., 2012; Tsuchiya et al., 2018). Winter time MSTIDs also tend to display an equatorwards velocity component. For example, Shiokawa et al. (2012) conducted a statistical study of night-time MSTIDs over Japan and found a preferential propagation direction to the South West, with velocities of between 50 and 100 ms$^{-1}$, and wavelengths of 100-300 km. Similar results were also reported by Frissell et al., (2014), where they isolated MSTID characteristics of auroral and lower atmospheric origin.

Several recent papers have utilised the observing capabilities of the LOw-Frequency-ARray (LOFAR: van Haarlem et al., 2013) to examine sub-structures internal to a TID, on a variety of scale sizes, and which demonstrate a variety of interesting radio scattering phenomena. Boyde et al., (2022) successfully refined a model from the 1980's (Meyer-Vernet et al., 1981) to reproduce spectral caustics generated by scattering from small-scale TIDs with wavelengths of < 30 km. Dorrian et al., (2023) examined the scattering characteristics and evolution of small-scale substructure, on scale sizes < 20 km, within an MSTID as it propagated over the British isles. In Fallows et al., (2020), observations of ionospheric scintillation in LOFAR were used to observe the transition of two separate TIDs at different altitudes crossing over each other, with one propagating at D-region altitudes and the other at F-region altitudes.

In these papers the scattering features are generally referred to as 'scintillation' or 'diffractive scintillation' if the individual scintilla are on time-scales of approximately 10-seconds or less and derive from plasma irregularities in the ionosphere which have scale sizes significantly smaller than the local Fresnel scale. Scattering features which are longer lived, typically on time scales of several minutes (e.g. Dorrian et al., 2023), require scale sizes which are larger than the Fresnel scale. In these instances the ionosphere is behaving less like a highly-structured plasma, with essentially randomised variations in refractive index, and more like a deformable refracting concave lens. Hence the terminology used for these larger features is 'refractive scattering' or 'refraction' instead of 'scintillation'.

In this paper we present LOFAR observations of refractive ionospheric lensing from structures generating entirely asymmetric QPS, which were embedded within an MSTID. The structures were most likely propagating through a regional sporadic-E layer. The MSTID is tracked for over 1200 km across Europe and is seen to contain many examples of QPS internal to itself over the entire distance for which it is tracked. To our knowledge there are few previous examples of a wholly asymmetric series of QPS which are observed to hold their form over such a large distance, and observed in broadband (24-64 MHz), in which radio frequency-dependent behaviour is seen clearly. The observations were made on 17 December 2018, in the middle of Northern hemisphere winter, and in the absence of any significant geomagnetic activity. In Section 2 we describe the observation method and instrumentation used. The observation results are presented in Section 3, and in Section 4 the results are discussed with reference to previous observations and numerical modelling of this phenomena.

## 2. Method & Instrumentation

Any exo-atmospheric radio signal received at the Earth's surface is a combination of the inherent properties of the signal and variations in amplitude and phase imposed upon that signal by passage through the structured plasma of the ionosphere. Such signal variations have been used for remote characterisation of ionospheric behaviour for many decades (e.g. Wild & Roberts, 1956; Bowman, 1981; Aarons, 1982). These signal variabilities can be broadly defined as refractive ionospheric focussing / defocussing (Koval et al., 2017; Koval et al., 2019; Boyde et al., 2022; Dorrian et al., 2023), in which the signal may vary steadily over several minutes, and much more rapid diffraction induced ionospheric scintillation, in which the signal variations occur on timescales of a few seconds or less (e.g. Singleton, 1974; Fejer & Kelley, 1980; Yeh & Liu, 1982; Tsunoda, 1988; Carrano et al., 2012). As pointed out by (Koval et al., 2017), under conditions of refractive ionospheric lensing at frequencies of <100 MHz, one may also observed fluctuations in the apparent spatial position of the radio source of up to several degrees, which may also contribute to signal strength variations. Refractive lensing by ionospheric irregularities, particularly in a radio astronomy context, was also investigated in earlier studies (Meyer-Vernet et al., 1981; Spoelstra & Kelder, 1984; Mercier, 1986; Mercier et al., 1989). Detailed reviews of the underlying physical mechanisms of ionospheric radio scattering are given by Booker (1981), Aarons (1982) and, more recently, Priyadarshi (2015).

Many previous ionospheric scintillation studies over recent decades have utilised signals from the global navigation satellite system (GNSS) network and ground based GNSS receivers to monitor TEC (strictly speaking, slant TEC) above the receiver. This technique has yielded a wealth of information about the behaviour of the ionosphere and is used in many space weather studies and forecasting models (e.g. Groves et al., 1997; Kintner et al., 2007; Wang et al., 2020; Elvidge & Angling, 2019; Otsuka et al., 2021), and is regularly applied in the monitoring and study of TIDs and small scale irregularities (e.g. recently Figueiredo et al., 2023; Borries et al., 2023; Prikryl et al., 2022; Ren et al., 2022; Themens et al., 2022; Zhang et al., 2022; Bolmgren et al., 2020; Zhang et al., 2019; Nykiel et al., 2017). GNSS satellites transmit encoded radio signals, typically at two or three frequencies in the Ultra High Frequency (UHF) band, which preserve the phase information of the signal at the top of the atmosphere and, enable phase delay induced by the ionosphere to be calculated and the path integrated plasma density to be determined.

LOFAR is a purpose-built radio telescope for observing natural radio sources which, like GNSS signals, are subject to ionospheric distortion. A key difference however is that natural radio sources are inherently broadband emitters, which enables ionospheric radio scattering features to be observed on many different frequencies simultaneously. What natural radio sources lack is any signal encoding which would enable one to ascertain phase delay after the radio waves have transited the ionosphere. None-the-less, broadband observations of ionospheric scintillation with LOFAR have yielded numerous results which would not have been possible using single or dual channel signals from artificial satellites. For example, as previously mentioned, Fallows et al., (2020) used LOFAR to detect and monitor the behaviour of two TIDs at different altitudes (one in the D-region, one in the F-region) as they passed across each other over Northern Europe. The presence of both plasma populations was revealed, in this instance, by the presence of two distinct scintillation arcs in the delay-Doppler spectra (2D Fourier transform) of the LOFAR dynamic spectra, which would not have been observable using single channel observations of the scintillation.

LOFAR is a networked array of, at the time of writing, 51 receiving stations spread across Europe, with most stations located in the Netherlands. Each LOFAR station can function as a radio telescope in its own right and is comprised of two arrays of antennas, the high-band antennas (HBA) which receives signals within the range 110-270 MHz, and the low-band antennas (LBA) which operates between 10-90 MHz. In this study we use data from the LBA arrays at each LOFAR station in the network, on 200 channels, each with a 195.3 kHz sub-band width at 10.5 ms time resolution. To exclude excessive RFI contamination, the LBA bandwidth used was restricted to 24.99-63.86 MHz. Each channel therefore produces a one-dimensional time series of received signal power from the observed radio sources. When combined in a single plot, the output from all channels forms the dynamic spectrum, an example of which is shown in Figure 2.

Fifty one LOFAR stations were used to simultaneously observe the strong radio sources Cygnus A (3C405: RA 19hr 59min 28s Dec. 40.73°) and Cassiopeia A (3C461: RA 23 hr 23 min 24s Dec. 58.82°), across all frequency channels, hereafter referred to as Cyg-A or Cass-A respectively. Of these stations, 24 are approximately co-spatial and located in the LOFAR core at 52.9°N, 6.9°E, and are labelled CS001-CS501. A further 14-stations are located throughout the Netherlands, labelled RS106-RS509, and are referred to as 'remote' stations. The remaining 13 stations are international, with the Easternmost station in Poland (PL612: 53.6°N, 20.6°E), and the Westernmost in the Republic of Ireland (IE613: 53.1°N 7.9°W). The most Northern and Southern stations used were in Sweden (SE607: 57.4°N 11.9°E) and France (FR606: 47.4°N, 2.2°E), respectively. The LOFAR field-of-view in this study thus encompassed approximately 28° of longitude and 10° of latitude centred on Northern Europe.

From the perspective of all LOFAR stations in the network, Cass-A is high circumpolar, whilst Cyg-A can drop to ~3° as seen from the LOFAR core. This viewing geometry thus enables long periods of continuous observations at a variety of elevations and azimuths. The wide-bandwidth, high-time resolution, and geographical spread of stations thus combine to make LOFAR an excellent and highly sensitive data source for ionospheric observations. The observing window for this study runs from 04:30:00-07:59:59.7 (all times henceforth are in UT) on the morning of 17-December 2018. The raw data for these observations may be accessed via the LOFAR long-term archive at https://lta.lofar.eu, using observation ID L691424 under project LT10_006.

## 3. Observations & Analysis

3.1 Geophysical context

Ground based magnetometer data from numerous stations in Northern Europe and Scandinavia recorded no geomagnetic activity from ~0000 on 17 December 2018 until a sub-storm late in the evening, well after the LOFAR observations had finished at 0800. From 0000-1200 on 17 December, the SMU and SML indices (Newell & Gjerloev, 2011) were <50 and featureless. The global Kp-index did not rise above 2 over the same time period, and there was only 1 small non-flaring active region (NOAA AR12731) approximately in the centre of the solar disk, as this was during the deep solar minimum at the end of Solar Cycle 24. No solar flares had been detected in the preceding 3-days and the interplanetary magnetic field at 1 AU was weak ($\leq$ 5nT) and Northwards oriented throughout. Geomagnetic conditions were therefore very quiet. Figure 1 shows the SMU / SML indices and the IMF clock angle for the period 0000-1200.

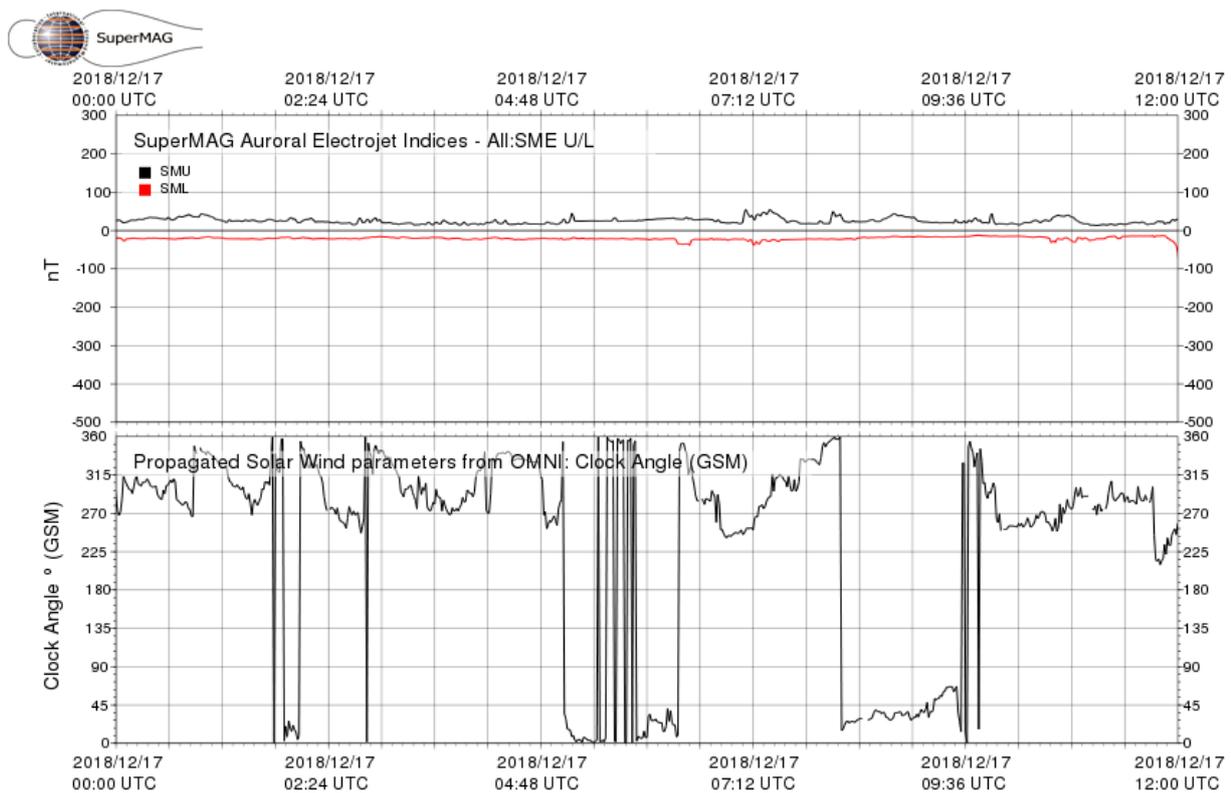

**Figure 1.** Top panel: SMU / SML indices or auroral electrojet activity from 0000-1200. Bottom panel: IMF clock angle over the same period showing mostly Northwards oriented IMF, especially during the LOFAR observing window from 0430-0759.

3.2 Dynamic Spectra.

RFI is present in some of the LOFAR data and so to mitigate this to an acceptable level each sub-band channel was median filtered using a sliding window of 50 data points which equates to approximately half a second. Data points exceeding 5σ above the standard deviation for the filtered data were removed. Some cases of RFI still remain after this process; where most of a given frequency channel is still contaminated then the data from that channel is excised. Furthermore, over the 3.5 hour observing window, both radio sources change significantly in elevation and azimuth and so each channel on all stations exhibits a source elevation dependency which was removed by fitting a third order polynomial, and then subtracting. The resulting data is then displayed as a mean-centred dynamic spectrum of normalized received signal power (see Figure 2 as an example), with observing frequency on the vertical axis, and time on horizontal axis.

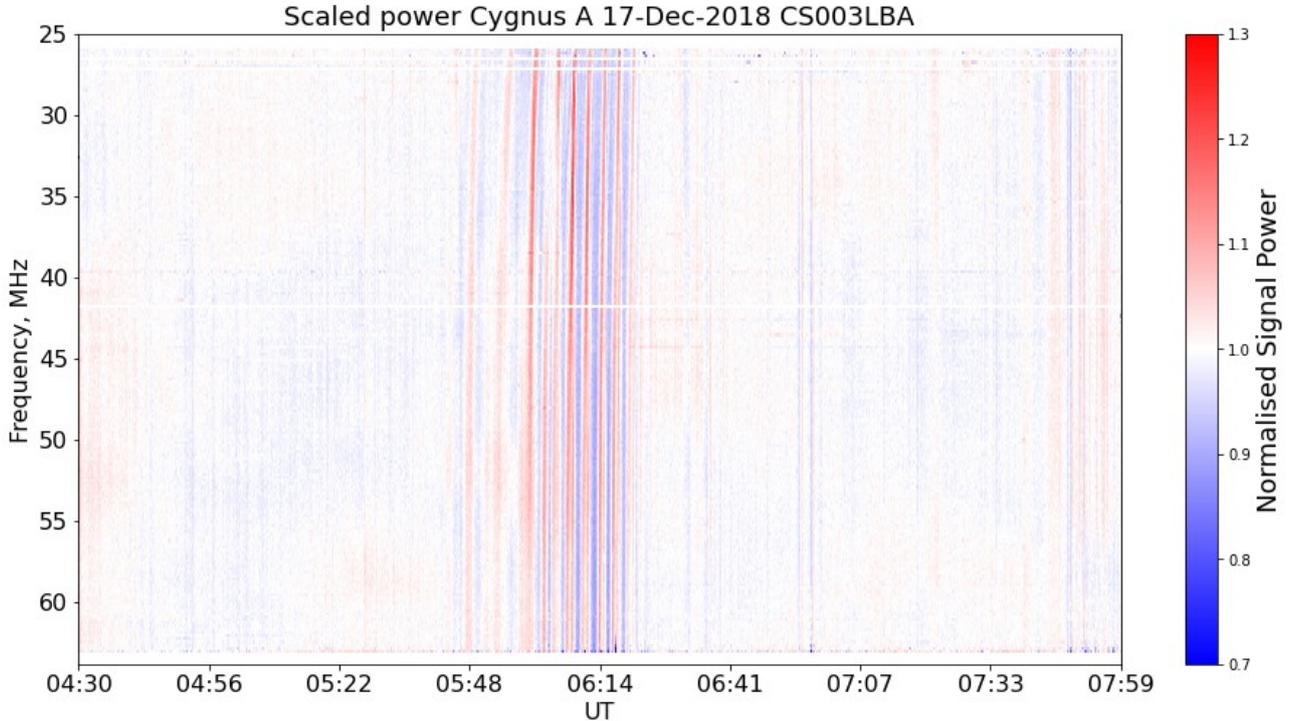

**Figure 2**. Example dynamic spectrum from the observation of Cyg-A using the LOFAR core station CS003 LBA covering the full observing window. The horizontal streaks are channels in which RFI contamination necessitated their excision. The vertical features at ~0600 are caused by ionospheric scattering. Signal power is scaled linearly, and shown by the colour bar.

The first signs of ionospheric activity were detected in Cass-A data from the Polish LOFAR station PL612LBA at 0503 on 17 December. Figure 3 (top panel) shows the dynamic spectra from PL612LBA observations of Cass-A. Until 0503, the normalized signal intensity was essentially flat and featureless, signifying an unperturbed ionosphere in the raypath. We then observe a rapid onset of an ionospheric perturbation which manifests as vertical curvilinear features spanning the full observation bandwidth, which begin with a broadband signal fade followed by a bright series of signal enhancements indicating rapid changes in plasma density along the raypath. The increasing drift of the features at the lower observing frequencies are due to increasingly strong ionospheric scattering, whereby the majority of the signal received at the station is from ionospheric structure which is off-axis from the centre of the station beam and manifests, in these data, as a 'delay' in arrival time when compared to the appearance time of the same feature at the higher frequencies.

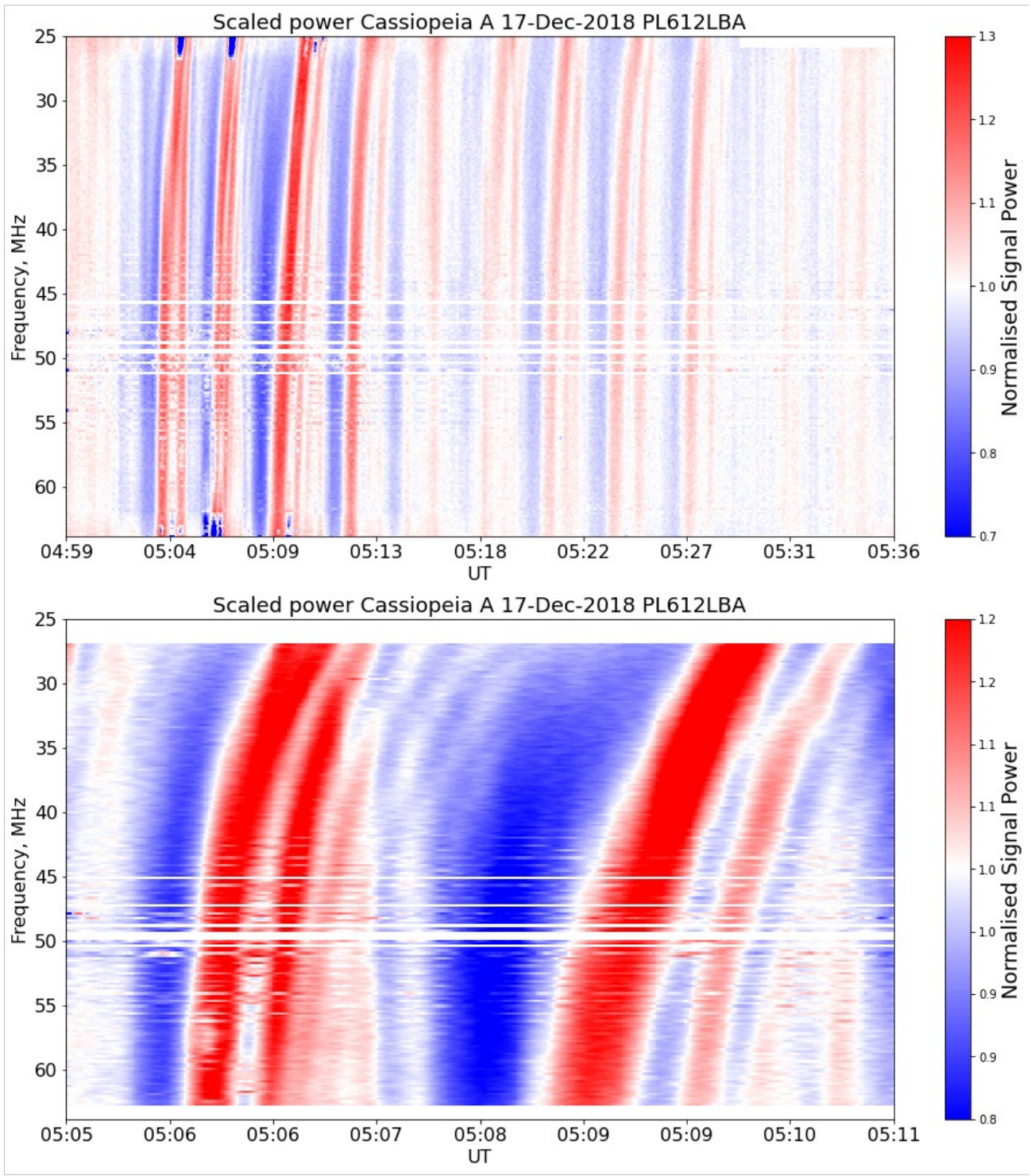

**Figure 3.** Dynamic spectra from the observations of Cass-A using Polish LOFAR station PL612LBA on 17 December 2018. Top panel shows perturbation from onset at 0503 to dissipation at 0527. Bottom panel shows a zoomed in and saturated view of two of the signal enhancements at 0506-0511, showing the repeated but progressively fainter fringes which follow the initial signal fade and initial brightening. The data in these plots are at 1-second time resolution. Horizontal streaks are excised RFI.

The features are quasi-periodic, with four of them seen clearly while at least another 5 fainter examples are observed following. On closer inspection (Figure 3, bottom panel) we can observe that each of the distinct signal fades and enhancements are followed by a series of further secondary signal enhancements and fades which are progressively fainter and shorter lived than the initial one.

After a few minutes, the process is repeated. The total time span for which the entire feature exists in the raypath at this LOFAR station is approximately 23-minutes, and extends from onset at 0503 to 0526 at which point the ionosphere returns to its previously unperturbed state. The horizontal streaks are excised RFI contamination.

Similar structures were then observed to appear time-sequentially in other LOFAR stations to the West-SouthWest, in Germany, then the Netherlands, before the final signatures of it were seen in the French and UK stations towards the end of the observing window. Figure 4 shows a selection of representative examples of dynamic spectra from different LOFAR stations across Europe. In each case where the perturbation was observed, a previously undisturbed ionosphere was seen to rapidly transition to a perturbed state, characterised by the sequence of signal enhancements and fades in the same manner as those observed in the Polish station. Different numbers of full-bandwidth signal fades/enhancements were observed in different stations, and on both radio sources, and in many cases the enhancements were immediately followed by a short series of secondary enhancements of decreasing intensity and time length. There was also considerable variation in the time over which the full event was seen from different LOFAR stations, and on each radio source. In some cases the event lasted ~10-minutes (e.g. RS307 on Cass-A), whilst in others it was seen for over an hour (e.g. DE604 on Cyg-A). Event onset and end times across the LOFAR network are given in Table 1.

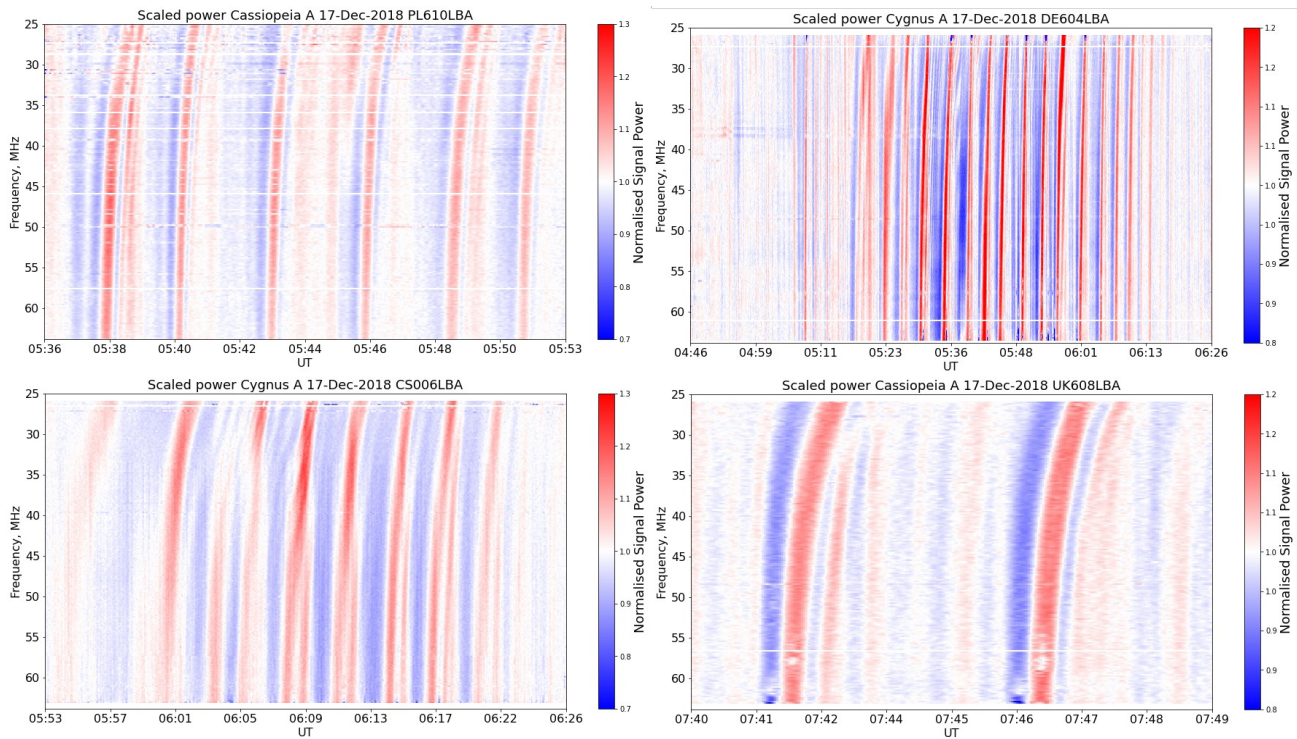

**Figure 4.** Examples of the perturbations detected across Northern Europe from geographically widely spaced LOFAR stations with PL610LBA (Poland) at top left showing at least 6 signal enhancements/fades, CS006LBA in the LOFAR core (Netherlands) at bottom left with 8, DE604LBA (Germany) showing at least 14 and UK608LBA (United Kingdom) showing 2. Zooming in on individual features show that many, from all stations where the detection was made, exhibited several secondary fade/enhancement fringes of decreasing intensity over typically <2 minutes (e.g. See Figure 3).

In the case of the German station DE603LBA, some 20 distinct enhancements were detected in Cyg-A observations, whereas in other stations (e.g. UK608LBA) only 2 or 3 were seen. In some cases the feature lay in the raypath for up to an hour whereas in others the duration was shorter. Data from DE603LBA and DE604LBA also exhibited the largest number of the secondary fainter fringes following the initial signal brightenings (Figure 5), in which up to 4 are seen to follow the initial one. In all cases where the feature is observed, the fainter secondary fringes always follow initial enhancements.

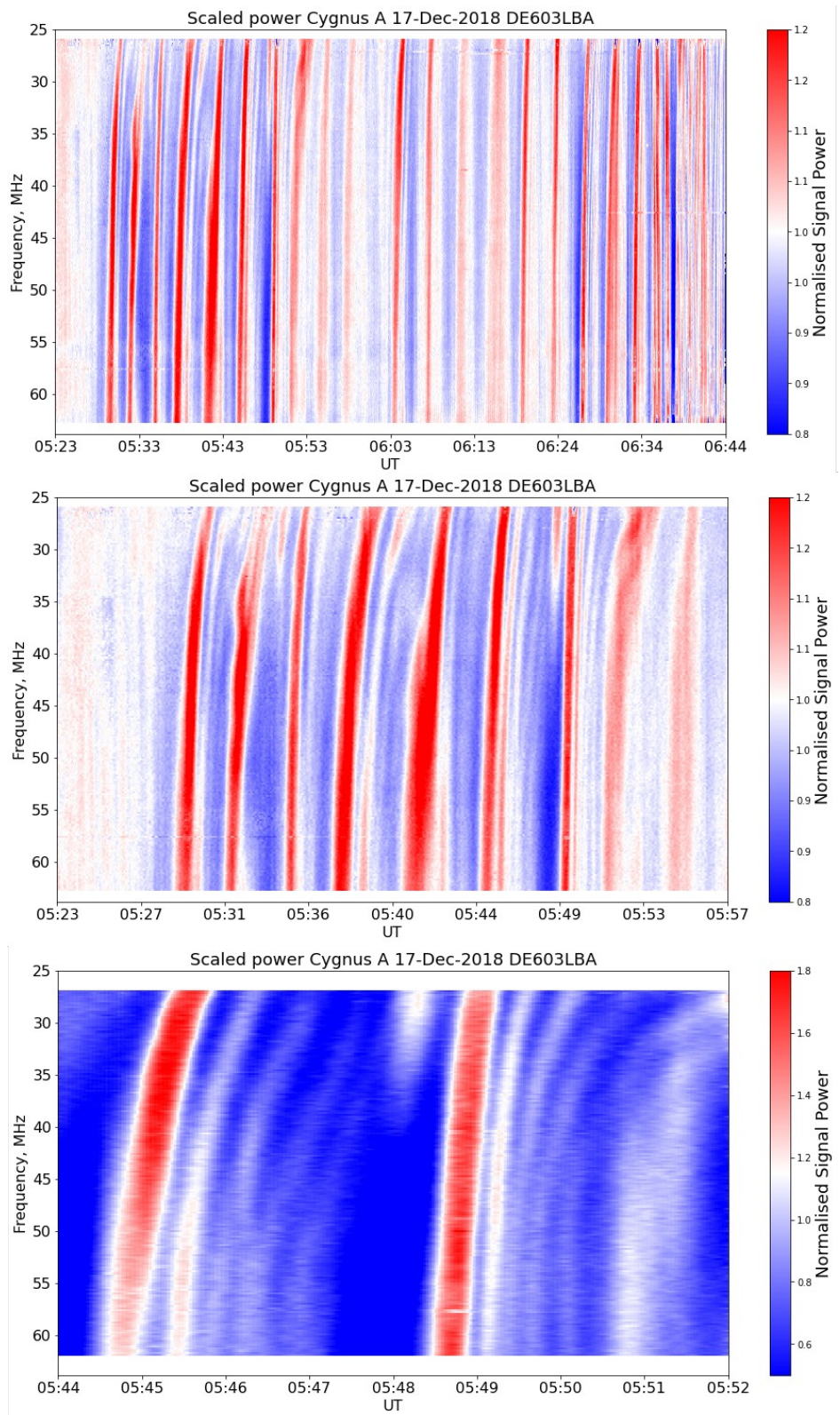

**Figure 5.** DE603LBA data from Cyg-A observations showing (top panel) the full TID extent which runs from 0530-0640 and exhibits at least 20 visible signal enhancements. A closer view of the brightest enhancements from 0528-0552 in which secondary fainter fringing can be seen clearly following initial enhancements is shown in the middle panel. A saturated view of the two enhancement features at 0544-0552 in which at least 4 shorter lived and progressively fainter secondary signal enhancements can be seen following the initial brightening (bottom).

In no cases was it observed that fainter secondary-type fringes preceding the primary examples. On some stations, such as SE603LBA in Sweden, the feature was not detected at all, whereas on others, including all three Polish LOFAR stations, it was seen in one radio source only (Cass-A) and not the other. This implies that the radio scattering plasma is indeed localised to the ionosphere and highly unlikely to be coming from more distant plasma motions such as in the solar wind where plasma feature scale sizes maybe of order $10^5$-$10^6$ km (e.g. Sheeley et al., 1997). Notice that in many cases these data sets are completely independent of each other with the two different radio sources being observed by different LOFAR stations.

In Figure 6 several examples of the secondary fringing are shown from different LOFAR stations across Europe. The colour scales have been offset from mean-centred to enhance the contrast and highlight the fringes. In some instances at least 6 secondary fringes are observed after the primary enhancement. The geographical separation of these observations and on different radio sources show that the plasma structure creating the QPS signal in LOFAR is holding its form over large distances. In all cases where it is observed, the secondary fringing follows and never precedes the primary signal enhancement. Each panel shows a 10-minutes of data and have the same normalized signal power scaling.

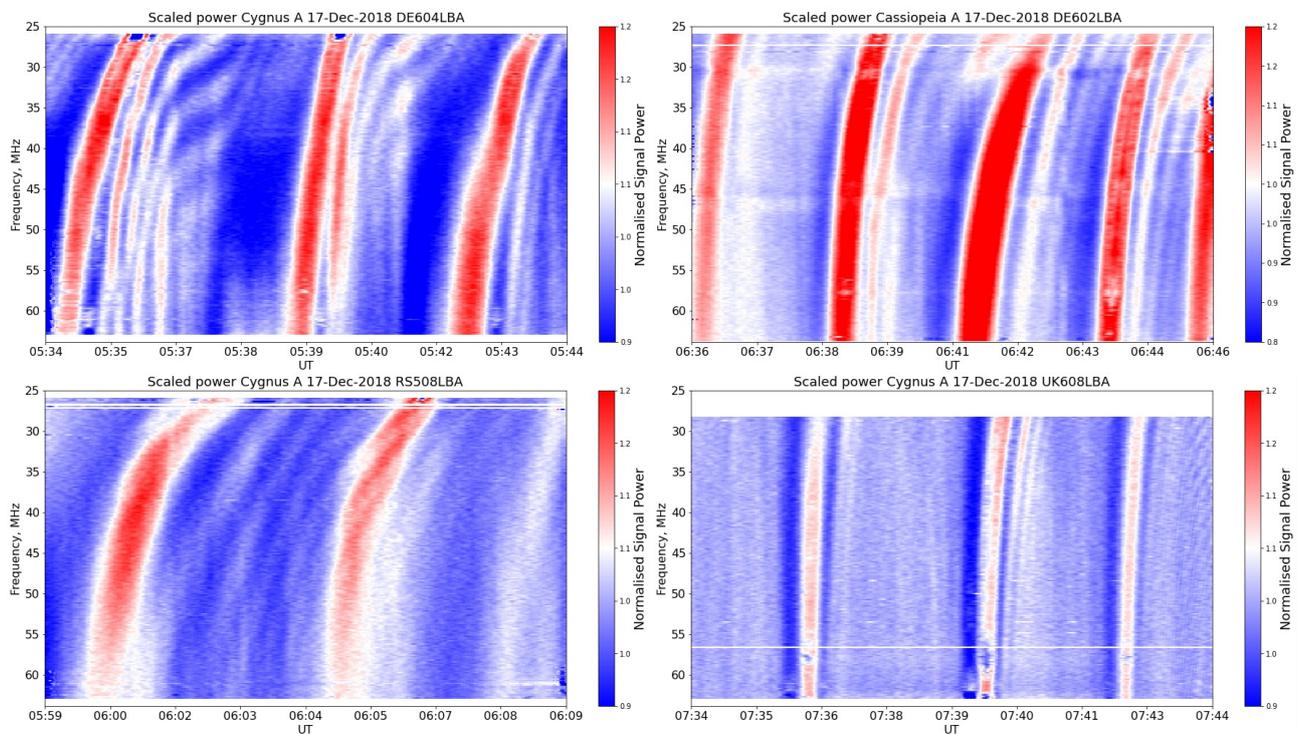

**Figure 6.** Examples of secondary fringing observed at LOFAR stations in Germany (top left, top right), the Netherlands (bottom left), and the UK (bottom right), on both radio sources. Colour scales are offset from mean-centred to enhance fringing contrast. In some instances up to 6 or 7 secondary fringes are detected whereas in others only 1 or 2 are seen. Each dynamic spectrum is of 10-minute duration and uses the same linear scaling for signal power.

Each dataset from each LOFAR station was examined and onset times identified visually. This was complicated somewhat by the fact that as the signal enhancements are curvilinear, the onset is not the same on all observation frequencies. Onset times are therefore given to the nearest minute and are shown with a representative sample of LOFAR stations and their approximate geographical locations in Table 1. Given the very close proximity of stations in the LOFAR core it was not possible to meaningfully separate feature onset times for each of these stations and so only one core station is shown. In cases where LOFAR stations are particularly close geographically (such as in the core), and thus have very densely clustered raypaths, only one LOFAR station from that area was included as it was not possible to clearly differentiate onset times.

| LOFAR Station | Coordinates | Cyg-A onset | Cyg-A end | Cass-A onset | Cass-A end |
|---|---|---|---|---|---|
| PL612LBA (Poland) | 20.5° E 53.6° N | Not seen | Not seen | 0504 | 0528 |
| PL610LBA (Poland) | 17.1° E 52.3° N | Not seen | Not seen | 0509 | 0531 |
| PL611LBA (Poland) | 20.5° E 50.0° N | Not seen | Not seen | 0509 | 0533 |
| DE604LBA (Eastern Germany) | 13.0° E 52.4° N | 0523 | 0610 | 0533 | 0550 |
| DE603LBA (mid-Germany) | 11.7° E 51.0° N | 0528 | 0638 | 0551 | 0630 |
| DE602LBA (South mid-Germany) | 11.3° E 48.5° N | Not seen | Not seen | 0618 | 0710 |
| DE605LBA (Western Germany) | 6.4° E 50.9° N | 0621 | 0658 | 0647 | 0655 |
| DE601BA (Western Germany) | 6.9° E 50.5° N | 0629 | 0705 | 0644 | 0654 |
| SE607LBA (Southern Sweden) | 11.9° E 57.4° N | Not seen | Not seen | Not seen | Not seen |
| RS208LBA (remote, Netherlands) | 6.9° E 52.9° N | 0557 | 0622 | 0630 | 0642 |
| RS307LBA (remote, Netherlands) | 6.7° E 52.8° N | 0601 | 0625 | 0635 | 0646 |
| RS310LBA (remote, Netherlands) | 6.1° E 52.8° N | 0603 | 0629 | 0641 | 0655 |
| CS005LBA (core, Netherlands) | 6.9° E 52.9° N | 0553 | 0622 | 0623 | 0639 |
| UK608LBA (Southern England) | 1.4° W 51.1° N | 0736 | 0742 | Not seen | Not seen |
| FR606LBA (North West France) | 2.2° E 47.4° N | Not seen | Not seen | 0756 | 0759 |
| IE613LBA (Ireland) | 7.9° W 53.1° N | Not seen | Not seen | Not seen | Not seen |

**Table 1.** Event onset and event end times (all UT) estimated by eye for a representative sample of LOFAR stations during the observing window, approximately in order of detection time.

3.3 Altitude

Before calculating accurate geographic positions of the feature it was first necessary to establish the altitude at which it propagated. As the feature was observed primarily over Northern Europe we examined data from the Juliusruh ionosonde and the co-located medium-frequency (MF) radar.

Figure 7 shows four ionograms from the nearby Juliusruh ionosonde covering approximately the same time period. The ionograms clearly show evidence for sporadic-E at an altitude of 110 km throughout the observing window. However, the sporadic-E is non-blanketing such that significant amounts of reflected echoes are also recorded from the F-region. This may be due to the sporadic-E having a more filamentary structure as opposed to a uniformly optically thick layer. F-region echoes in these ionograms are particularly useful however as they show no evidence of significant spread-F or trace bifurcation which one would typically associate with the passage of TID perturbations in the field-of-view (Bowman, 1990). The F-region appears undisturbed throughout the observing window, while the presence of a filamentary or granular, non-blanketing and long lived sporadic-E layer over the same time period is more consistent with the TID travelling at this altitude (110 km).

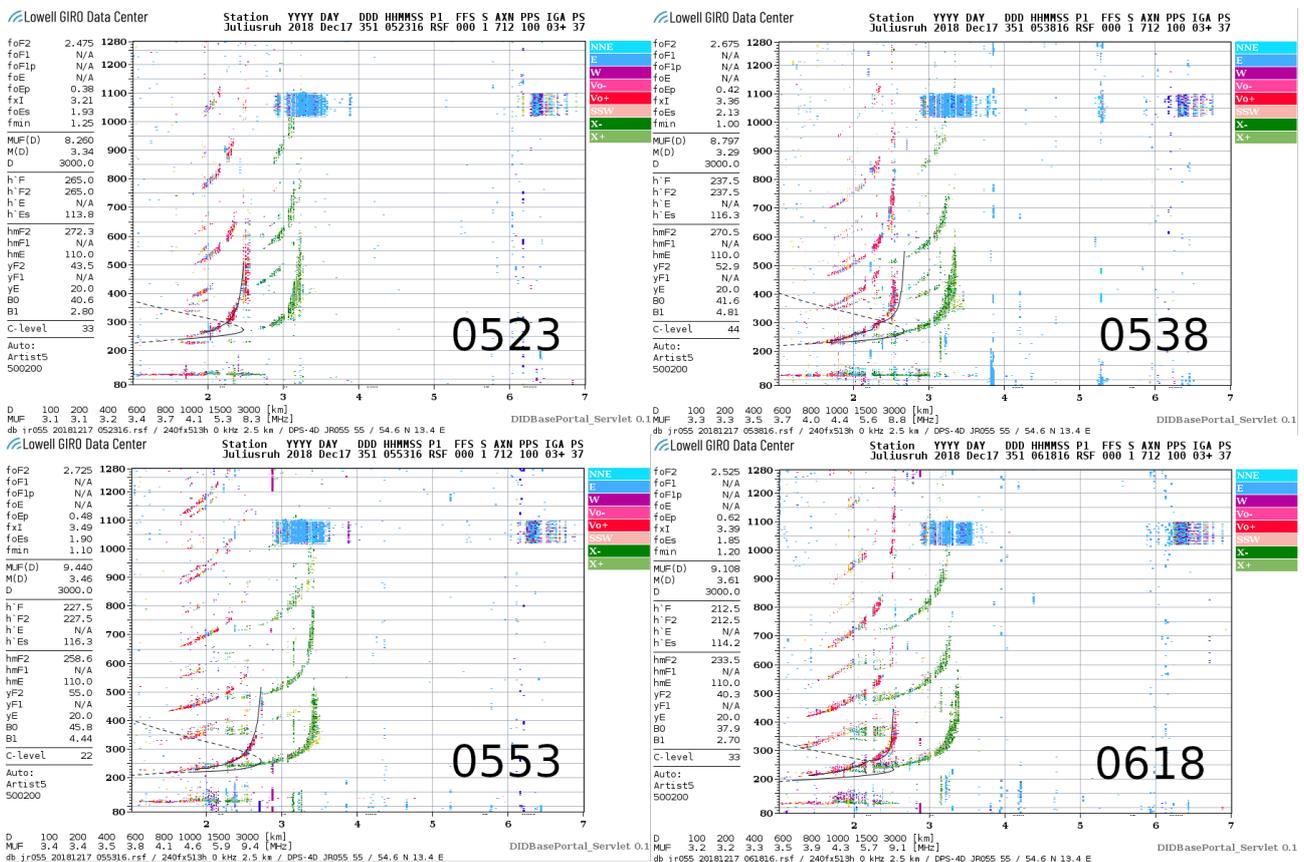

**Figure 7.** Four ionograms from the Juliusruh ionosonde (13.37 W, 54.63 N), with the times in UT shown in large text. All four exhibit non-blanketing sporadic-E at 110 km, notably the same altitude as the structure observed in the MF radar. Backscatter echoes from the F-region are also visible overlapping in frequency with the sporadic-E and, throughout, show little evidence of any F-region disturbance.

The Juliusruh MF radar consists of a Mills-Cross-Antenna with 13 crossed half wave dipoles operating at a peak power of 64 kW. It has an altitude resolution of 1 km and operates at a frequency of 3.18 MHz, or 94 m in wavelength. Figure 8 shows Juliusruh MF radar data from the day of the observation. The top left panel shows backscatter from E-region features over 24-hours on 17 December 2018. The feature of interest for this study is shown in the zoomed in view in the top right panel, and is visible between 0500-0600, coinciding with the presence of the disturbance in the LOFAR observations from the German stations DE603 and DE604. It is extended in altitude, and centred at approximately 110 km. A smaller altitude extended feature at similar altitudes is also seen shortly before 0700. The larger bottom panel is an angle-of-arrival (AOA) plot covering the period 0502-0556. The x- and y- axis are the directional cosines (E/W and N/S respectively). The MF radar data has a time resolution of 3-minutes; each point in the AOA plot depicts the mean AOA position, from each 1 km range bin in a given 3-minute time window, with colour representing altitude.

There are several points to note here. Firstly, the MF radar was pointed at zenith throughout the observations. At an altitude of 115 km, the width of the radar's main beam is approximately 35 km. The radar is most sensitive to structures with gradients in refractive index of approximately half the transmission wavelength. So given the wavelength is 94 m, the mean structure sizes to which the radar is most sensitive are ~47 m in size. Secondly, the altitudes are variable but clustered around 110 km, indicating the presence of structured plasma rather than a thin uniform plasma layer at constant altitude. Thirdly, all the echo positions are oriented approximately NW-SE. Finally, it is worth considering that the phase velocity of a propagating radio signal which encounters a structured plasma will decrease, prolonging the echo return time and hence giving an overestimation of range. It is possible therefore that the altitude of the higher points in the AOA plot are somewhat overestimated.

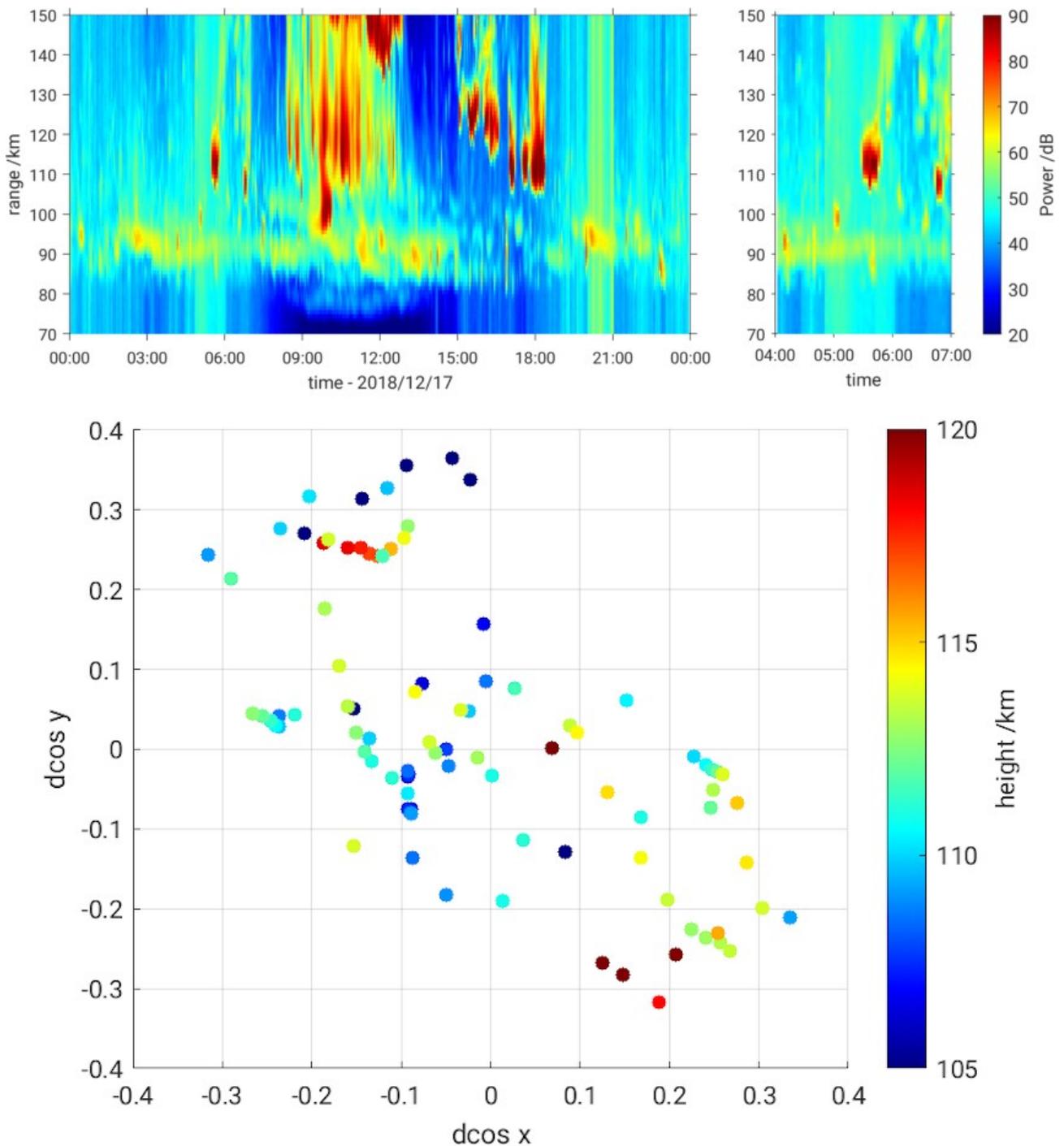

**Figure 8**. MF radar plot from the Juliusruh radar, 17 December 2018 (top left panel), showing a distinct structure in the E-region, extended in thickness over ~10 km and centred at an altitude of 110 km, between 0500-0600 (top right panel). An angle-of-arrival plot is shown in the bottom panel with each point representing the mean AOA position from every 1 km range bin in each 3-minute time bin covering the period 0502-0556. The colours of each point show the altitude as indicated; the width of the radar's main beam at 115 km altitude is approximately 35 km.

3.4 Propagation

It is evident from the onset times in Table 1 that the feature was propagating approximately Westwards. Given that not all stations detected it, the feature was therefore also localised to within a restricted geographic extent of the ionosphere roughly encompassing the Baltic Sea, Southern Scandinavia and the Northern coasts of Germany, Denmark, and the Netherlands. The method for calculating the positions of the ionospheric pierce point (IPP) for a given observation described in Section 3 of Dorrian et al., (2023) is dependent on establishing the altitude of an ionospheric feature of interest. Using the height of the sporadic-E layer established in the previous section, several maps are presented (Figure 9) showing the geographic positions of the IPPs. In all cases the IPPs are projected to an altitude of 110 km and rotate clockwise along their respective arcs throughout the observing window from 0430-0759. The pale orange arcs show the position of the IPP for Cyg-A, with the dark orange shaded part of the arc showing the proportion which was occupied by the feature. Feature onset, where seen, is shown with a yellow spot and accompanying UT time. Pink arcs show the equivalent for Cass-A, with red shading indicating the proportion of the arc in which the event was seen and, again, onset time and location for the event are indicated. The position of the Juliusruh ionosonde and MF radar, from which E-layer height was ascertained, is also shown, as is the nearby GNSS station used later in Section 3.5. The maps have been split to show different regions of Europe and a few LOFAR stations each so as to avoid over cluttering the plots. Note that the feature was not detected in raypaths which covered central/Southern Germany and Western Poland nor Southern Sweden, thus constraining the propagation to approximately over the Baltic Sea and the Northern coasts of Germany, Denmark, and the Netherlands.

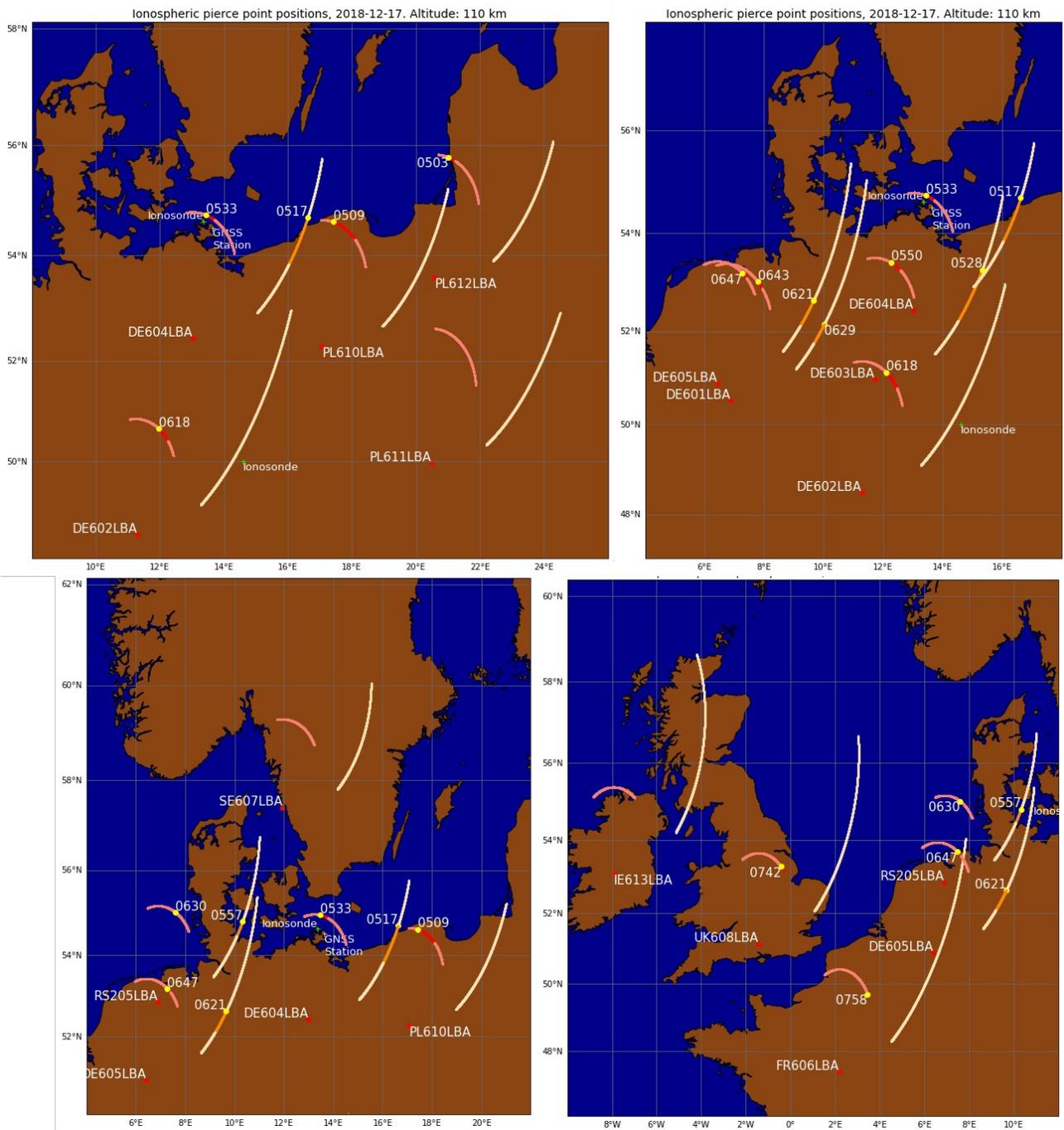

**Figure 9.** IPP arcs for selected LOFAR stations from East to West Europe at a projected altitude of 110 km for the duration of the observing window (0430-0759). All IPPs rotate clockwise along their respective arcs. Pale orange arcs are for Cyg-A with dark orange shading showing the proportion of the arc over which the event was observed and the onset time of the event indicated. In arcs where the event was not seen there is no shading or start times shown, which helps to constrain the geographic extent of the feature. Pink arcs are the IPPs for Cass-A with dark red shading showing the extent of the event and accompanying onset times. Positions of the Juliusruh ionosonde and MF radar are shown by green spots, as is the nearby GNSS receiver station. To avoid over cluttering the maps, different regions of coverage are shown in the different panels. Note the absence of a detection in the Swedish station (SE607, bottom left panel) and several of the arcs in Eastern Poland (top left panel) which offer some constraints on the geographic extent of the feature.

In Figure 10 an overlay of several dynamic spectra is shown from different LOFAR stations at different times in the observing window and their corresponding IPP arcs projected to 110 km. In all cases where observed, feature onset could be established visually given the very quiet background ionosphere which preceded and followed the passage of the feature through the raypath. This behaviour is consistent with the passage of a TID through the region of interest, with the QPS indicating the presence of sub-structure within an individual TID wave. This is discussed further in section 3.5.

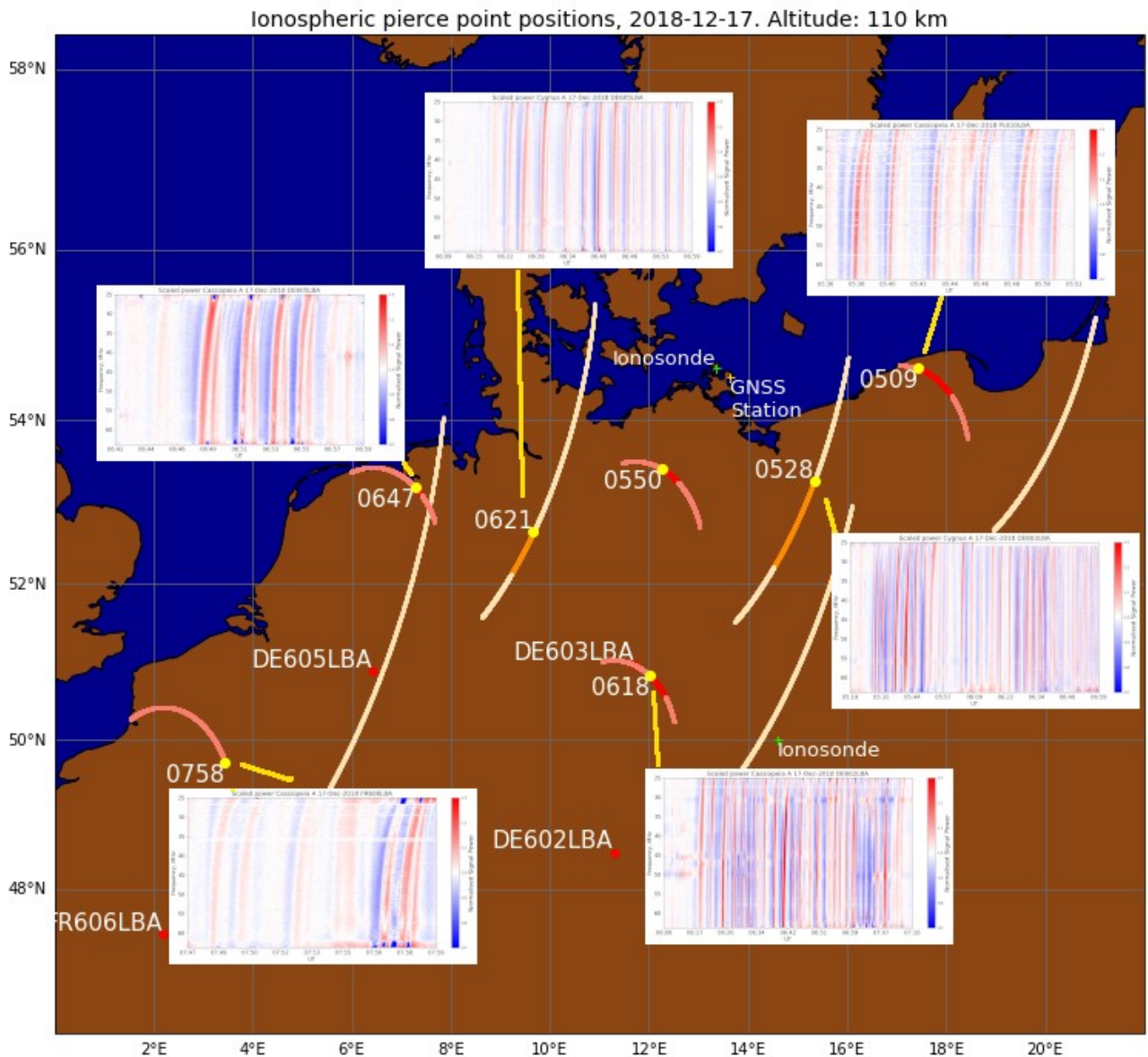

**Figure 10.** Several dynamic spectra with their associated IPP arcs at 110 km altitude for selected LOFAR stations in the network.

Calculating the propagation characteristics was performed by using the known onset times and estimated coordinates of the TID from the LOFAR data as shown in Figures 9 and 10. A modelled plane wave was propagated from an origin point using the coordinates for the first IPP in which the feature was detected in Cass-A on PL612 (Figure 9, top left panel). Figure 11 shows a schematic for estimating wave propagation for a given azimuth and velocity over the spherical surface of the Earth.

A modelled TID wave is propagated from the first IPP (IPP$_1$) at a given velocity and azimuth, and the time taken to reach IPP$_2$´ is then estimated using Napier's rules for right-angled spherical triangles. The objective is to calculate the distance from IPP$_1$ to IPP$_2$´ which can in turn be used to calculate the expected arrival of the modelled plane wave at IPP$_2$. The angle θ is the difference between the known azimuth of IPP$_2$ from IPP$_1$ and the azimuth of propagation of the modelled plane wave. Given that IPP$_1$ and IPP$_2$ both lie on the same plane through the centre of the sphere, then the distance between them (λ) is simply the Great Circle distance. This distance is then equivalent to a distance angle (φ) along the Great Circle given by:

$$\phi = \left(\frac{\lambda}{2\pi r_E}\right) 2\pi$$

The distance angle IPP$_1$-IPP$_2$´ (φ´) can then be calculated by:

$$\phi' = \tan^{-1}(\cos(\theta)\tan(\phi))$$

The actual distance from IPP$_1$ to IPP$_2$´ is then given by:

$$IPP_1 IPP_2 ' distance = \left(\frac{\phi'}{2\pi}\right) 2\pi r_E$$

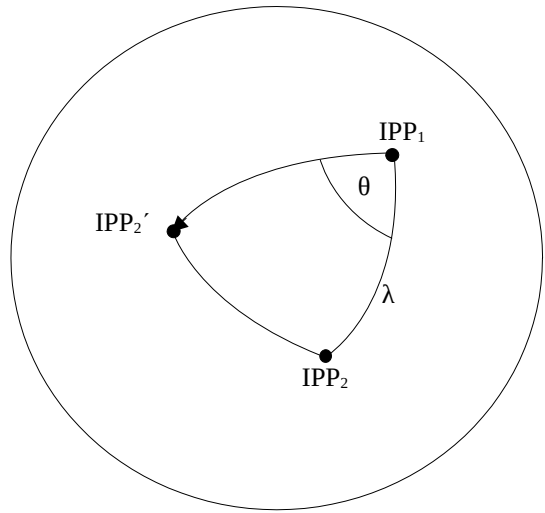

**Figure 11.** Schematic for estimating the arrival time of a modelled plane wave at IPP$_2$, using IPP$_1$ as the origin. The TID wave is propagated at a given azimuth and velocity from an origin point IPP$_1$ (arrowed line) and is assumed to arrive at IPP$_2$ and IPP$_2$´ at the same time given that it is being modelled as a plane wave.

The propagation of the TID can then be estimated by projecting a variety of modelled plane waves with different azimuths and velocities until the best fit is found, based on the minimum average residual minutes between the estimated arrival of the TID at each IPP with the actual observed arrival from the LOFAR data. The results of this analysis, as seen in Figure 12, yielded a most likely TID propagation azimuth of 255° at a velocity of 170 ms$^{-1}$, which is quite consistent with the time-sequential appearance of the TID in LOFAR station data shown in the maps above (Figures 9, 10).
This approach relies on a few assumptions; namely that the event seen in each LOFAR dataset is the same one, that it does indeed propagate as a plane wave in a direction perpendicular to the wave front, that onset can be unambiguously identified by eye from background noise in a given dynamic spectrum, and that the TID velocity remains constant throughout. It is also likely that errors will become more apparent with greater distances and times from the origin as any deviation from a strict plane wave propagation will become more apparent, and that Earth is an oblate spheroid rather than a perfect sphere. Even with these considerations, the modelled wave properties yields a good fit when comparing expected onset times from the modelled wave with actual onset times seen in the LOFAR data, with most points in Figure 12 lying within a few minutes of the perfect fit line.

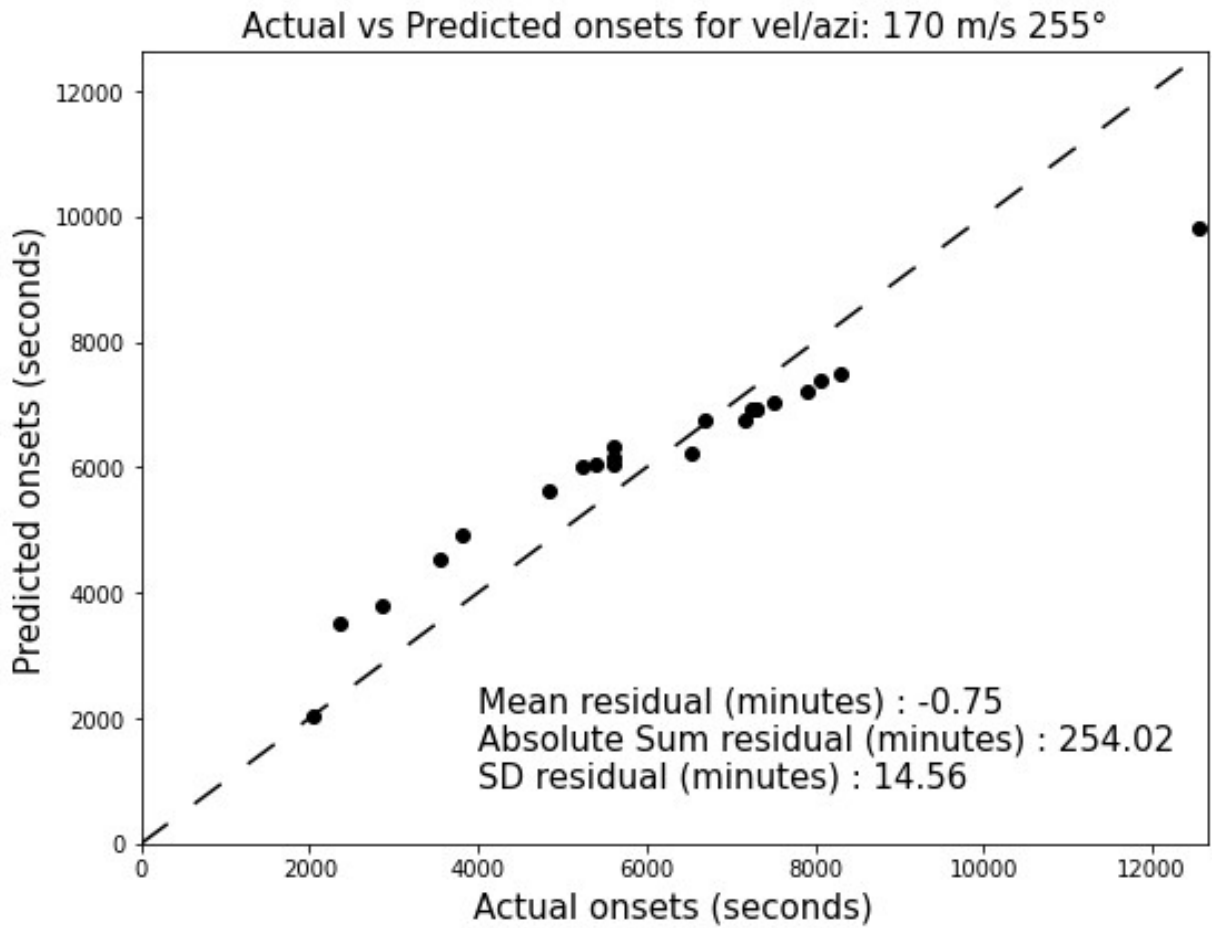

**Figure 12.** Actual vs predicted onsets for a TID modelled as a plane wave propagating from the origin point at the first IPP in which it was detected, through all of the other IPPs at all LOFAR stations used in the analysis. In the event of LOFAR stations being very closely space geographically, only one is chosen from that location given that little difference in onset times can be detected.

3.5 GNSS Data

To attempt to characterize the full morphology of the TID, TEC anomaly maps of the region of interest were produced using the methods described in Themens et al., (2022). The TEC anomaly maps time-increment by 2-minutes and show enhancements and depletions of plasma density of < 0.02 TECu. Each grid cell is 0.25° of latitude/longitude. Each map shows variations in plasma density with respect to a 30-minute subtracted average background. Areas of no detectable vTEC variation or areas with no data due to low satellite elevations or limited lines of sight over sea are shown in white. A video showing ionospheric activity from 0000-1200 on 17 December 2018 is included in the supplementary information to this paper. Presented in Figure 13 are several frames coinciding with the LOFAR observations of the QPS over Germany at approximately 0600. The most outstanding features are a series of distinct plasma density enhancements and depletions with amplitudes no greater than +/- 0.02 TECu, appearing as red and blue latitudinally extended wave structures, which propagate West-South West in good agreement with the TID propagation azimuth from LOFAR data in Section 3.4.

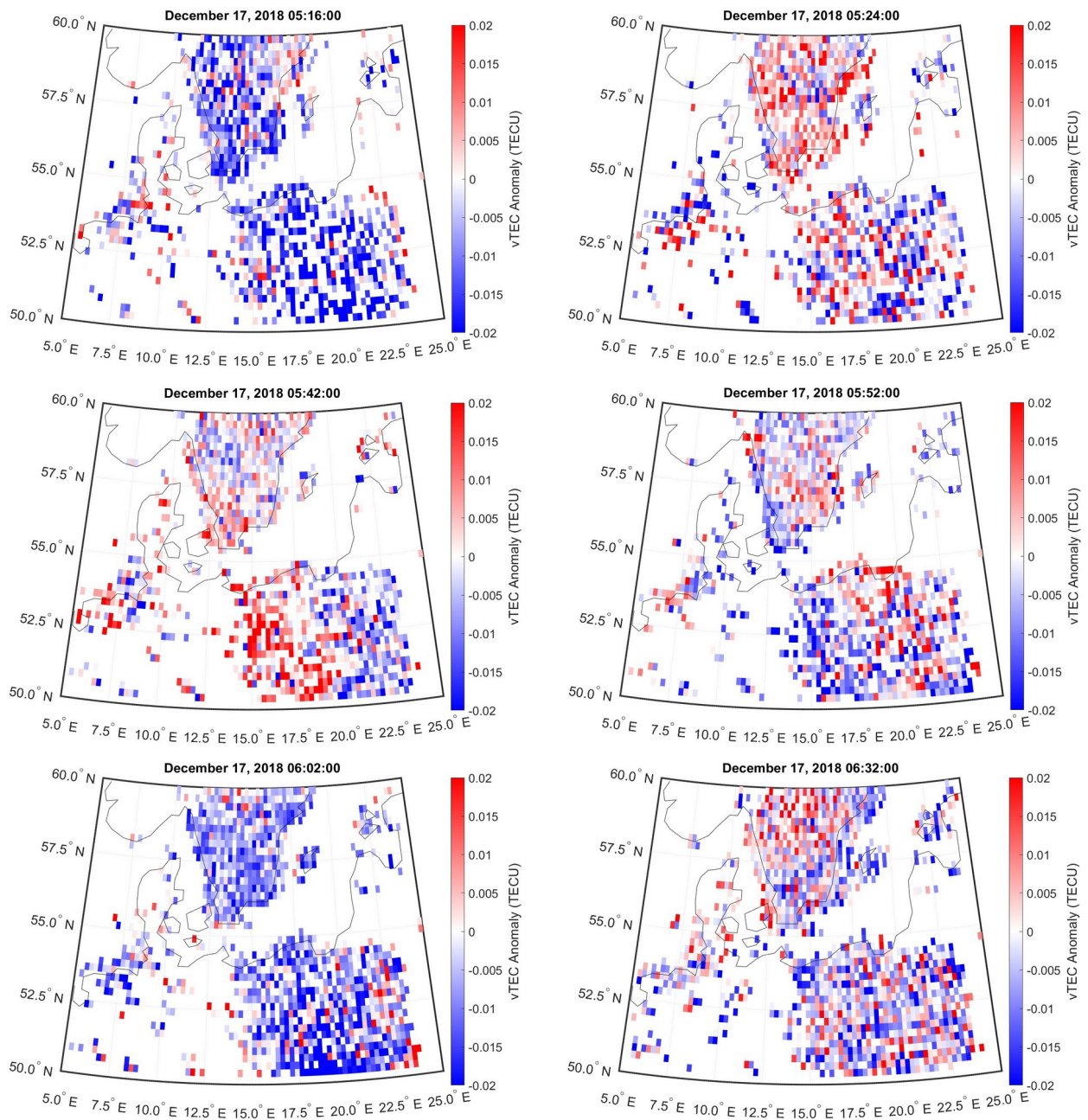

**Figure 13.** 30-minute averaged background subtracted vTEC maps with times in UT shown in bold text. TID propagation is seen covering the Baltic Sea and Northern Germany at the time of the LOFAR observations as consecutive blue and red wave-structures showing sequential plasma density enhancements and depletions. A video covering the period 0000-1200 UT covering the same area is provided in the supplementary information to this paper. The West-South Westerly propagation of the TID is in good agreement with estimated TID propagation from using event onset times in the LOFAR data and the orientation of the strongest echoes in the MF radar data (Figure 8).

In addition to producing vTEC anomaly maps one can also examine structures observed in single-satellite GNSS data in instances where the line-of-sight to the satellite is proximate to nearby LOFAR lines-of-sight. Given that the most detailed LOFAR observations of QPS were seen in data from the German LOFAR stations DE603 and DE604 (Figures 4-6), and that we also used data from the Juliusruh ionosonde and MF radar, vTEC anomaly signatures in GNSS data from the Juliusruh GNSS station were also examined.

Visibility to 31 GNSS satellites was available from the Juliusruh GNSS station on this day and the IPP for the lines-of-sight to at least one satellite (PRN 11) passed very close to LOFAR lines-of-sight from DE603 and DE604, projected to an IPP altitude of 110 km. In Figure 14 the LOFAR IPPs for DE603 and DE604 at an IPP of 110 km altitude, for both radio sources (Cass-A in orange, Cyg-A in cyan), are shown on the same map as the IPP arc for GNSS satellite PRN 11 (in yellow) as seen from the GNSS station near Juliusruh. The position of the GNSS station itself is also shown. The satellite was visible at Juliusruh from 0438-0755 respectively, which was very close to the LOFAR observing window of 0430-0759. Given the position and propagation of the TID ascertained from analyses in the previous sections, it was highly likely the GNSS IPP would intersect with the TID. Also shown in the bottom right panel of Figure 14 are the Rate of TEC index (ROTI) for satellites PRN 11 and 14, which show small-scale ROTI of <= 0.05 TECu along their IPP arcs.

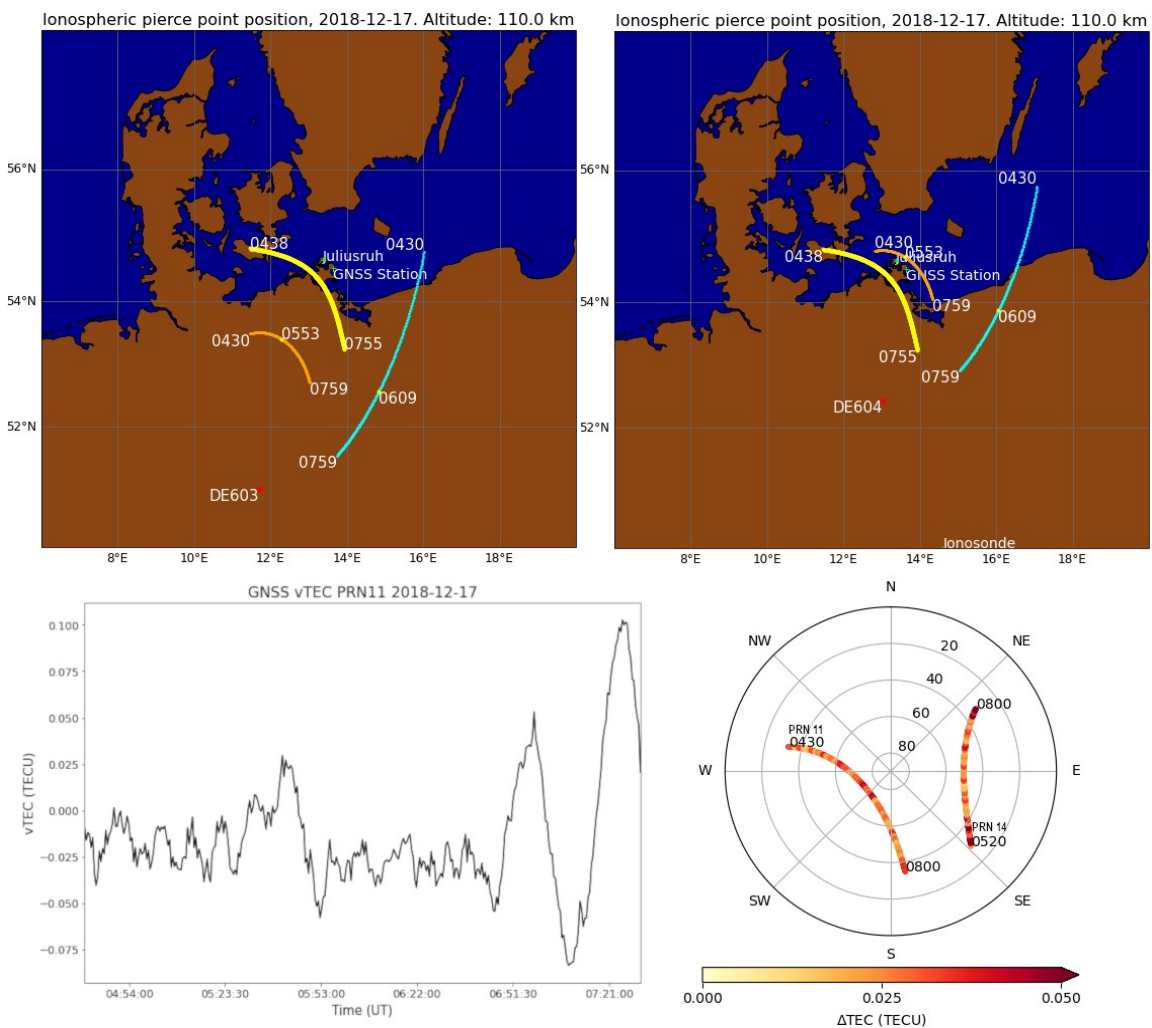

**Figure 14.** Top right, top left: LOFAR raypaths for stations DE603 and DE604 showing the IPP for each radio source projected to an altitude of 110 km (orange arcs are Cass-A, cyan arcs Cyg-A). The IPP arc, also at 110 km altitude, for GNSS satellite PRN 11 is shown in yellow. The time-points near the middle of each LOFAR IPP arc show the approximate mid-point of the event for that observation. Bottom left: vTEC data from the Juliusruh GNSS station (ground position shown in maps) for satellite PRN 11 covering approximately the same time window as the LOFAR observations. Bottom right: Sky-view from the Juliusruh GNSS receiver station to PRN 11 and, at slightly greater distance from the LOFAR IPP arcs, PRN 14, showing ROTI along the lines-of-sight.

A series of small amplitude (< 0.05 TECu) and short period waves are the dominant feature in the vTEC data from PRN 11, from 0438 to 0651 UT as the IPP passes through the TID, followed by larger amplitude variations at the end of the visibility window. Given that the raypaths from the LOFAR stations and the GNSS station to PRN 11 are not identical it is unlikely that exactly the same wave structures are being sampled in each; the data are presented here as an approximate comparison. In Figure 15 (top panel) the vTEC data from PRN11 in the period dominated by highly regular small amplitude waves (~0430-0630 UT) is shown, with one data point every 30-seconds. The vTEC data was smoothed using a third-order polynomial Savitzky-Golay filter (Savitsky & Golay, 1964), with a window size of 21 data points, and a periodogram using Welch's method was then produced (bottom panel), which revealed a clear peak in signal power at a period of ~280 seconds.

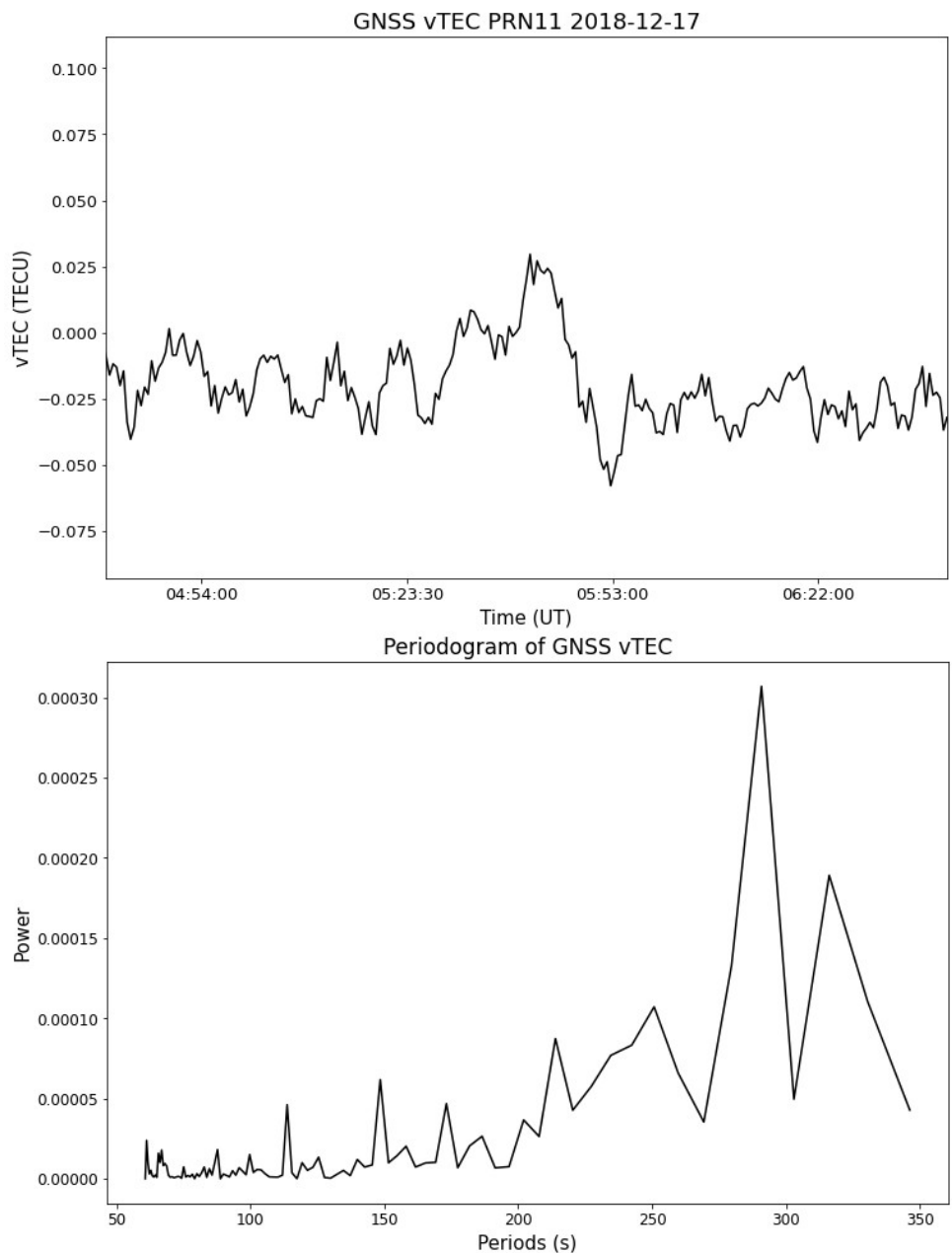

**Figure 15.** Top panel: 30-minute de-trended vTEC anomaly data from GNSS PRN11. Bottom panel: Welch's method periodogram showing peaks in signal power between 250-320 seconds.

3.6 LOFAR periodicities

The periodicities of the QPS were broadly consistent across all LOFAR datasets and representative examples are shown in the 2-D periodograms in Figure 16. The periodograms were constructed by performing Welch's method on each frequency channel for a given dynamic spectrum. In all cases periodicity of the major sub-structure signal enhancements across all stations, and on both radio sources, were between 150-250 seconds. In some cases the periodograms show singular peaks across all wavelengths, whereas in others, finer structure is visible at shorter periodicities which is due to clearer definition of secondary fringing patterns in those datasets. At propagation velocities of ~170 ms$^{-1}$, these periodicities equate to distances of approximately 25-43 km between each significant signal power enhancement and the next, which is much shorter than wavelengths typical of an MSTID. The Fresnel scale for a given observing wavelength is given by $F_D = \sqrt{2\lambda L}$, where $\lambda$ is wavelength and L is the distance from the LOFAR station to the scattering screen. A frequency of 40 MHz corresponds to a wavelength of 6.7 m. If L is the altitude of the scattering plasma, in this case 110 km, then $F_D$ is ~1.2 km. Scale sizes of 25-43 km are therefore characteristic examples of ionospheric lensing from substructure within the TID itself (e.g. Koval et al., 2017) given that they are much larger than the Fresnel scale at the observation frequencies used.

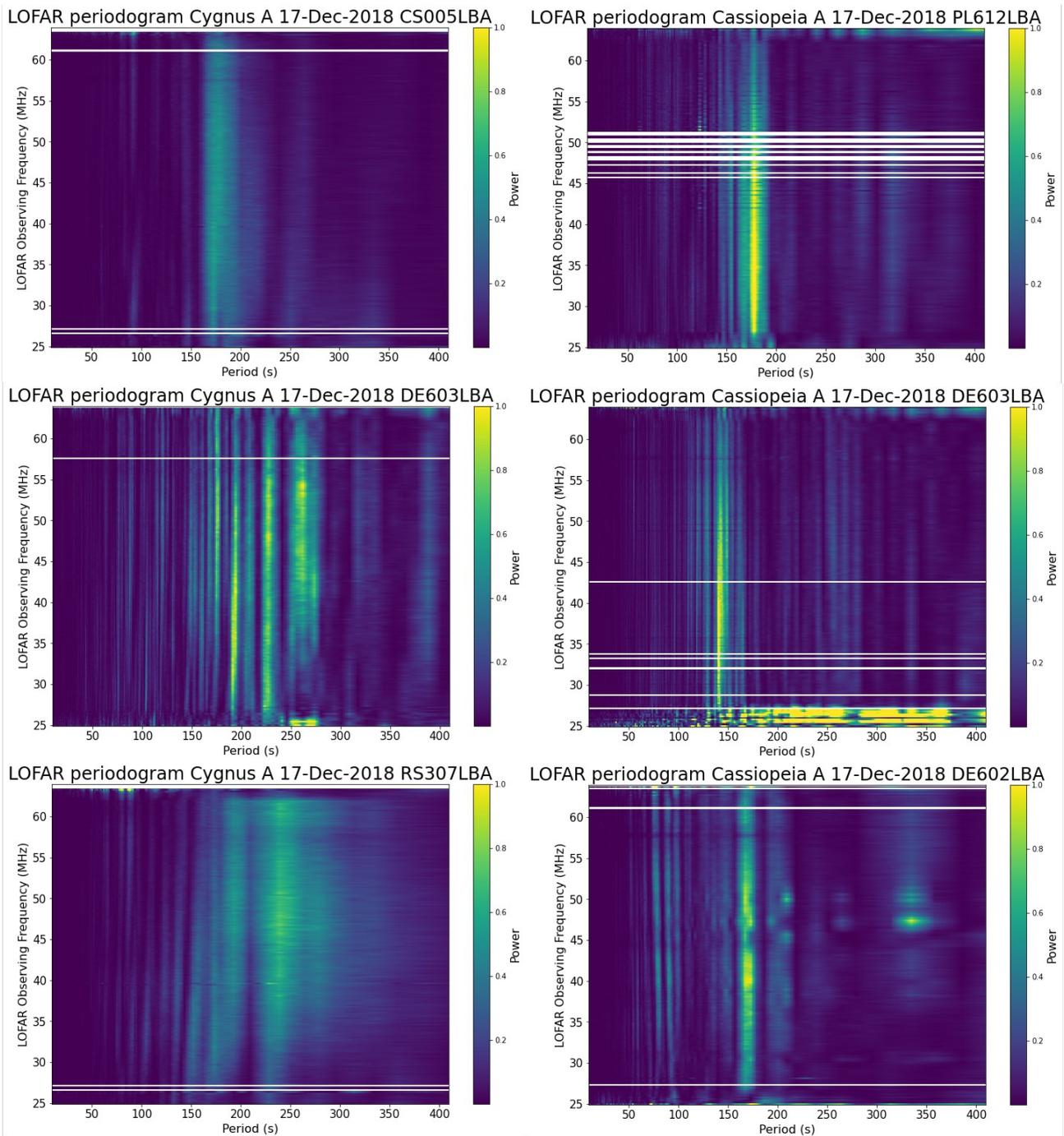

**Figure 16.** Representative 2-D periodograms of dynamic spectra from several LOFAR stations showing consistent dominant periodicities across all observation frequencies of between 150-250 seconds on both radio sources. The left column of panels are produced from Cass-A observations and the right column from Cyg-A observations. Instances of greater definition of secondary fringes appear as multiple discrete peaks in periodicities (e.g middle left), whereas weaker fringing results in a smearing of the peaks over a greater width (e.g. bottom left). All panels are presented with the same time and signal power scales.

## 4. Discussion

The LOFAR signal patterns shown in this paper are consistent with examples of Type 1 QPS, as identified by Maruyama (1991), namely asymmetric and characterised by an initial signal fade and enhancement, followed by a series of decreasing intensity secondary fringes. These secondary fringes are referred to by Maruyama as a 'ringing pattern', akin to the damped oscillations of a bell.

The QPS signals are generated by plasma structures embedded within a SW propagating TID in a long-lasting sporadic-E layer at ~110 km altitude.

Modelling work by Maruyama (1995) and Bowman (1989) attributes the ringing pattern to interference fringes caused by Fresnel diffraction as the incoming radio waves traverse a narrow region of steep plasma density gradient on the trailing edge of disk shaped plasma irregularities which are most likely located in the E-region. In the examples presented here the amplitude of the primary signal fade is comparable to the amplitude of the primary signal enhancements. It is possible that the QPS arises from coupling between plasma populations moving along inclined field lines between the E- and F-regions. However, the absence of any strong evidence of disturbances to the F-region in the ionograms imply that the plasma structures causing the QPS were wholly contained within E-region altitudes. Furthermore, the MF radar AOA plot (Figure 8, bottom panel) clearly shows a NW-SE orientation of structured plasma which was clustered around 110 km in altitude. This is consistent with the wavefront of the TID extending perpendicular to the propagation direction (SW), and provides a very useful direct measurement of the altitude of the TID which is independent of LOFAR, the GNSS data, and the ionosondes.

Some previous authors have commented on the field-aligned nature of QPS features like these (e.g. Yamamoto et al., 1991). However, in the present study the azimuth of the LOFAR lines-of-sight to Cass-A varied from -5° to +25°, and from +34° to +67° on Cyg-A throughout the observing window. Yet the QPS is observed to hold its form and is detected by many LOFAR stations on both radio sources over several hours. During this time the TID passes through regions of different magnetic declination and inclination angles, and LOFAR lines-of-sight, all with different directions. In the region of its first detection in the Polish LOFAR stations, declination angle for 17 December 2018 is ~9° and over the North Sea declination angle is ~0°. Hence it seems unlikely that the plasma structures causing this QPS are field-aligned as they are observable at many different angles with respect to the local field. It should be noted that at the time of detection there was almost no geomagnetic activity, so the above estimates, which are based on IGRF-13, are likely to be quite reliable.

The appearance of the QPS in different LOFAR station data varies primarily in the number of individual waves observed. In the Polish LOFAR stations, about 5 or 6 were seen. Across Germany, in particular LOFAR stations DE602, DE603, and DE604, as many as 20 were seen. Towards the end of the observing window, the UK station sees only 3. This may be a function both of the evolution of the TID as it propagates and that different LOFAR lines-of-sight were intersecting the plasma in different locations and observing geometries. Generally speaking the QPS were most clearly observed over northern Germany, however the secondary ringing patterns were observable in many stations all across Europe.

The periodicities calculated from the LOFAR data range from ~150-250 seconds and the periodicity in the GNSS data from PRN 11 is ~280 seconds. The best estimate of TID propagation, as calculated using onset times for the QPS in consecutive LOFAR stations, yielded a velocity of 170 ms$^{-1}$ with an azimuth of 255°. The scale size of the plasma structures which produce the QPS can therefore be estimated using the periodograms, however, all of these periodicities will be subject to Doppler effects in varying degrees as the IPP for the LOFAR stations and the satellite move through the atmosphere at some fraction of the propagation velocity of the TID. Furthermore, the direction of movement of all these lines-of-sight vary considerably throughout the observing period as the IPPs trace out arcs of movement (see Figures 9, 10). Therefore some component of a given IPP velocity at a given point in time will map onto the estimated propagation of the TID. A further complication is that in order to calculate the periodograms, one needs to use accumulated LOFAR/GNSS data over a large section of the total observing window, thus also encompassing potentially large changes in the component of the IPP movement, which is mapped onto the TID movement. This is likely to smear out the Doppler effect to some degree.

Figure 17 shows the azimuths and velocities of the moving lines-of-sight to Cass-A and Cyg-A from LOFAR station DE604, and the same parameters for GNSS satellite PRN 11 from the Juliusruh GNSS station. The fastest moving line-of-sight was for Cyg-A which, near the beginning of the observing window, was moving at nearly 50 ms$^{-1}$. If the maximum velocity of these lines-of-sight were aligned either directly parallel with or anti-parallel to the propagation of the TID (170 ms$^{-1}$, 255°) then it could shift the calculated periodicities by a maximum of +/- ~30%. However, this velocity, and that of the GNSS satellite line-of-sight, both drop rapidly whilst the Cass-A lines-of-sight movement remains at ~10 ms$^{-1}$ throughout. Therefore the maximum possible error in the periodicities of 150-250 s (for LOFAR) and 280 s (for GNSS) is +/- 50-75 seconds, and +/- 93 seconds, respectively. It should be noted that these values are the worst possible cases, and that for most of the observation period, the IPP velocities through the atmosphere are much lower than this. Especially, as one can see from the azimuths, that the movement of the IPPs will be typically be off-parallel or off anti-parallel with respect to the propagation of the TID. In the case of Cass-A, the velocity of the line-of-sight is never more than 10% of the TID velocity, so the periodicities calculated for LOFAR data on this source should be quite accurate.

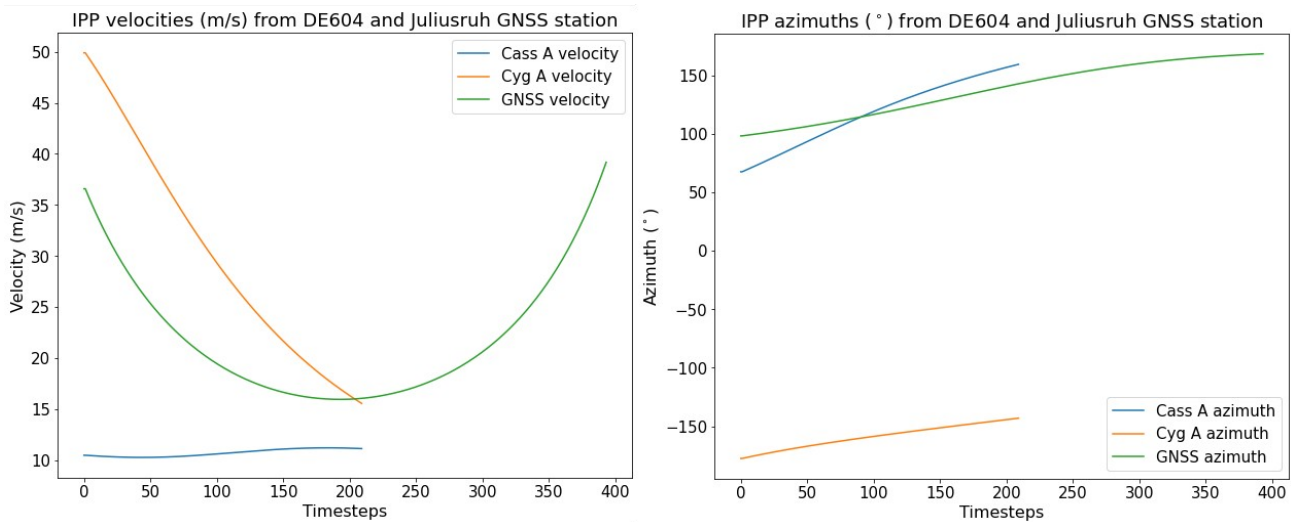

**Figure 17.** The left panel shows the velocities of the LOFAR lines-of-sight to Cass-A and Cyg-A, respectively, through the atmosphere at an altitude of 110 km and the same information for GNSS satellite PRN 11, as seen from the GNSS ground station near Juliusruh. The panel on the right shows the azimuth of movement of these lines-of-sight. The LOFAR lines-of-sight positions are calculated for every 60-seconds, while the GNSS satellite line-of-sight position is calculated for every 30-seconds, hence the greater number of times steps.

If the TID propagation velocity estimated from the LOFAR onset timings is reasonably accurate then the periodicities observed in LOFAR and from the GNSS satellite would yield plasma scale sizes of between approximately 25-43 km. As the periodicities are dominated by the initial deep signal fade and enhancement in each observed QPS, rather than the fringes, this would imply that each structure causing a QPS was between 20-40 km away from the next one, assuming the Doppler uncertainties outlined above are not significant.

That the TID in which the QPS were detected was observed over such a large distance is testament to the wide geographical spread of the LOFAR stations and highlights their utility as a wide field-of-view ionospheric observatory. The QPS were observed at the very start and end of the full observing window, so we cannot ascertain the full distance over which the TID travelled or where and when it may have dissipated. This contrasts with Dorrian et al., (2023) in which a TID was observed to break up and evolve over the distance for which it was observed.

Given that the TID here was observed all the way from over the North-East Europe to the North Sea, a distance of some 1200 km, it is not unreasonable to assume that its propagation prior to first detection in the Polish LOFAR stations could have carried it all the way from the auroral region, most likely somewhere over Northern Russia. High latitude magnetometer data from the region however showed very little activity in the preceding 5-hours before the start of the LOFAR observing window. Figure 18 shows SuperMAG (Gjerloev et al., 2012) ground based magnetometer plots from Kharasavey (66.8°E,71.2°N) and Bear Island (19.2°E, 74.5°N) from 2300-0900 on 17/18 December. Unfortunately the FUV instrument on the IMAGE satellite was not available at this time, so establishing a visual position for the auroral oval is somewhat inhibited. The absence of any significant auroral activity leaves open the possibility that this TID was generated from a terrestrial rather than space weather source, but a detailed identification of the initial driving source is not within the scope of this paper.

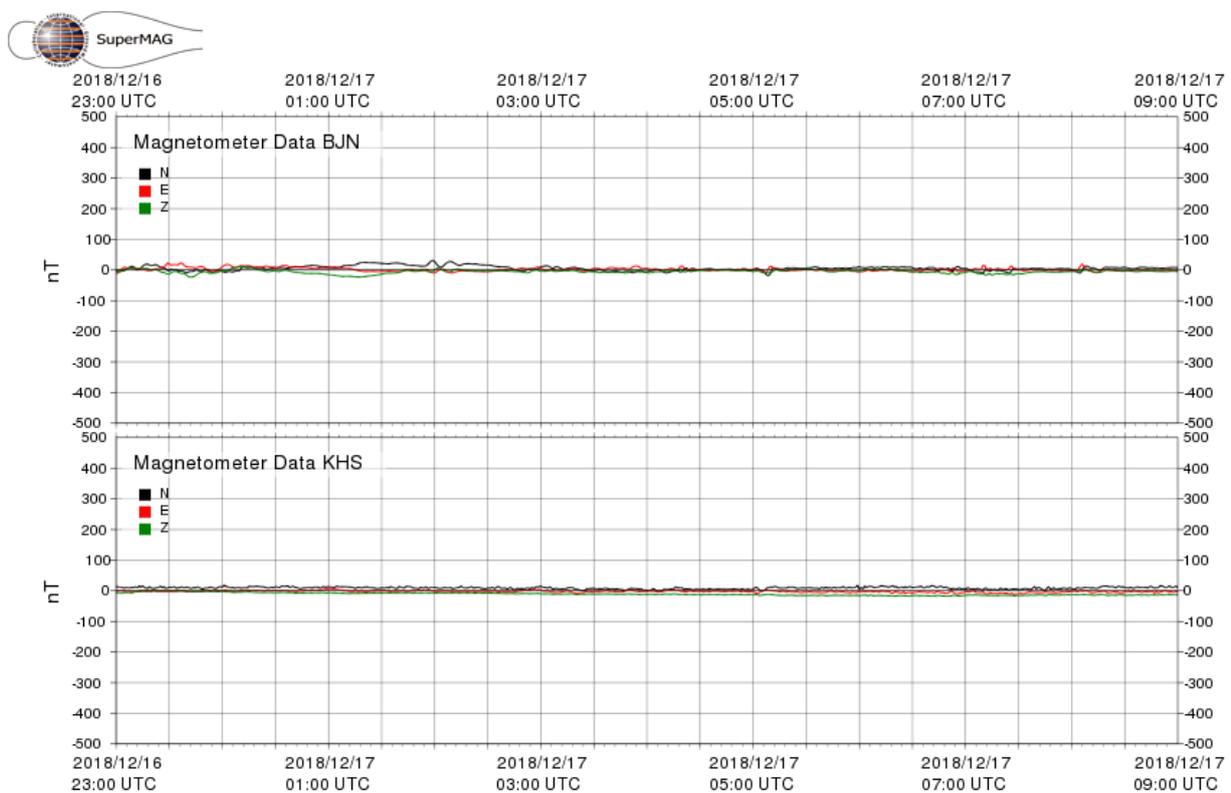

**Figure 18.** SuperMAG ground based magnetometer plots from 2300-0900 UT, 17-18 December 2018. The LOFAR observing window begins at 0430 UT on 18 December, and the first detection of the QPS in the Polish stations occurs at 0504 UT.

It is evident from the GNSS data that the magnitudes of the plasma structures generating the QPS were very small, being no larger than +/- 0.02 TECu. Future modelling work will be needed to establish whether these QPS generating structures are inherently long-lasting and hence preserve information about the driving processes of the TID, or whether they may arise continuously and spontaneously within a TID if the local ionospheric conditions permit it.

## 5. Conclusions

We have used the international Low Frequency Array to track the propagation of plasma structures embedded within a TID, generating type 1 asymmetric quasi-periodic scintillations over a distance of at least 1200 km in the mid-latitude ionosphere. The QPS are observed across the full longitudinal distance of the LOFAR network which imply that the plasma structures generating them held their form consistently over this distance. Information from the ionograms and the propagation altitude, which was directly measured by MF radar as 110 km, are consistent with a non-blanketing sporadic E-region, which remained in place for the full duration of the TID passage over Northern Germany. To our knowledge these measurements are the first broadband ionospheric scintillation observations of such a phenomena. Clear observing frequency dependent behaviour can be observed in the QPS, with the lower observing frequencies detecting the QPS up to 2-3 minutes after it was seen at higher frequencies due to strong ionospheric signal scattering at these lower frequencies. Contemporary GNSS data showed the large scale structure and propagation of the TID which were broadly in agreement with the propagation characteristics ascertained from LOFAR.

## Acknowledgments & Data Access

The LOFAR data used for this study can be accessed on the LOFAR long-term archive under project LT10_006 at https://lta.lofar.eu/. The International LOFAR Telescope (van Haarlem et al., 2013) is designed and constructed by ASTRON. It has observing, data processing, and data storage facilities in several countries, that are owned by various parties (each with their own funding sources), and that are collectively operated by the ILT foundation under a joint scientific policy. The ILT resources have benefited from the following recent major funding sources: CNRS-INSU, Observatoire de Paris and Université d'Orléans, France; BMBF, MIWF-NRW, MPG, Germany; Science Foundation Ireland (SFI), Department of Business, Enterprise and Innovation (DBEI), Ireland; NWO, The Netherlands; The Science and Technology Facilities Council, UK; Ministry of Science and Higher Education, Poland.

For the ground magnetometer data we gratefully acknowledge: INTERMAGNET, Alan Thomson; CARISMA, PI Ian Mann; CANMOS, Geomagnetism Unit of the Geological Survey of Canada; The S-RAMP Database, PI K. Yumoto and Dr. K. Shiokawa; The SPIDR database; AARI, PI Oleg Troshichev; The MACCS program, PI M. Engebretson; GIMA; MEASURE, UCLA IGPP and Florida Institute of Technology; SAMBA, PI Eftyhia Zesta; 210 Chain, PI K. Yumoto; SAMNET, PI Farideh Honary; IMAGE, PI Liisa Juusola; Finnish Meteorological Institute, PI Liisa Juusola; Sodankylä Geophysical Observatory, PI Tero Raita; UiT the Arctic University of Norway, Tromsø Geophysical Observatory, PI Magnar G. Johnsen; GFZ German Research Centre For Geosciences, PI Jürgen Matzka; Institute of Geophysics, Polish Academy of Sciences, PI Anne Neska and Jan Reda; Polar Geophysical Institute, PI Alexander Yahnin and Yarolav Sakharov; Geological Survey of Sweden, PI Gerhard Schwarz; Swedish Institute of Space Physics, PI Masatoshi Yamauchi; AUTUMN, PI Martin Connors; DTU Space, Thom Edwards and PI Anna Willer; South Pole and McMurdo Magnetometer, PI's Louis J. Lanzarotti and Alan T. Weatherwax; ICESTAR; RAPIDMAG; British Artarctic Survey; McMac, PI Dr. Peter Chi; BGS, PI Dr. Susan Macmillan; Pushkov Institute of Terrestrial Magnetism, Ionosphere and Radio Wave Propagation (IZMIRAN); MFGI, PI B. Heilig; Institute of Geophysics, Polish Academy of Sciences, PI Anne Neska and Jan Reda; University of L'Aquila, PI M. Vellante; BCMT, V. Lesur and A. Chambodut; Data obtained in cooperation with Geoscience Australia, PI Andrew Lewis; AALPIP, co-PIs Bob Clauer and Michael Hartinger; MagStar, PI Jennifer Gannon; SuperMAG, PI Jesper W. Gjerloev; Data obtained in cooperation with the Australian Bureau of Meteorology, PI Richard Marshall.

G. Dorrian, A. Wood, & B. Boyde were supported in this work by a research grant from the Leverhulme Trust (RPG-2020-140). H. Trigg was supported by a student bursary from the Royal

Astronomical Society (PI: G. Dorrian). DRT's contribution to this work is supported in part through the United Kingdom Natural Environment Research Council (NERC) EISCAT3D: Fine-scale structuring, scintillation, and electrodynamics (FINESSE) (NE/W003147/1) and Drivers and Impacts of Ionospheric Variability with EISCAT-3D (DRIIVE) (NE/W003368/1) projects. RAF was partially supported by the LOFAR4SW project, funded by the European Community's Horizon 2020 Program H2020 INFRADEV-2017-1 under Grant 777442.